\documentclass[12pt]{article}
\newif\ifclean
\cleantrue 
\usepackage[letterpaper]{geometry}
\usepackage{subfig}
\usepackage{amssymb,amsfonts}
\usepackage{graphicx}
\usepackage{hyperref}
\usepackage{cite}
\graphicspath{{{./}}}

\usepackage{xspace} 
\usepackage{amsmath,amssymb,amsthm}
\usepackage{bm}
\usepackage{xcolor}
\usepackage[normalem]{ulem}
\newcommand{\COMMENT}[1]{\textcolor{cyan}{{[ \sc{#1} ]}}} 

\newcommand{\red}[1]{\textcolor{red}{{#1}}}

\newlength{\figwidth}
\newlength{\figwidthtwo}
\newlength{\figwidththree}
\newcommand{\aref}[1]{App.\,\ref{#1}}
\newcommand{\fref}[1]{Fig.\,\ref{#1}}

\newcommand{\eref}[1]{Eq.\,(\ref{#1})}

\newcommand{\sref}[1]{Sec.\!~\ref{#1}}

\newcommand{\cref}[1]{Ref.\,\cite{#1}}
\newcommand{\crefs}[1]{Refs.\,\cite{#1}}

\newcommand{\ie}{{\it i.e.}\! }
\newcommand{\eg}{{\it e.g.}\! }

\newcommand{\etal}{{\it et al.}\! }

\newcommand{\Ac}{\mathcal{A}}
\newcommand{\Bc}{\mathcal{B}}
\newcommand{\Ic}{\mathcal{I}}

\newcommand{\Oc}{\mathcal{O}}

\newcommand{\Cbb}{\mathbb{C}}

\newcommand{\epsilonb}{\boldsymbol{\epsilon}}
\newcommand{\phib}{\boldsymbol{\phi}}

\newcommand{\ab}{\mathbf{a}}
\newcommand{\bb}{\mathbf{b}}
\newcommand{\eb}{\mathbf{e}}
\newcommand{\fb}{\mathbf{f}}
\newcommand{\yb}{\mathbf{y}}
\newcommand{\hb}{\mathbf{h}}

\renewcommand{\sb}{\mathbf{s}}

\newcommand{\xb}{\mathbf{x}}
\newcommand{\vb}{\mathbf{v}}

\newcommand{\Bb}{\mathbf{B}}

\newcommand{\Db}{\mathbf{D}}

\newcommand{\Fb}{\mathbf{F}}

\newcommand{\Eb}{\mathbf{E}}

\newcommand{\Sb}{\mathbf{S}}
\newcommand{\Ib}{\mathbf{I}}
\newcommand{\Rb}{\mathbf{R}}

\newcommand{\Tb}{\mathbf{T}}

\newcommand{\Xb}{\mathbf{X}}

\newcommand{\Wb}{\mathbf{W}}

\newcommand{\chib}{\boldsymbol{\chi}}
\newcommand{\varphib}{\boldsymbol{\varphi}}

\newcommand{\spn}{\operatorname{span}}

\newcommand{\tr}{\operatorname{tr}}

\newcommand{\dev}{\operatorname{dev}}
\newcommand{\sym}{\operatorname{sym}}

\newcommand{\partialb}{\boldsymbol{\partial}}
\newcommand{\delt}{\Delta\!\,t}
\newcommand{\grad}{\boldsymbol{\nabla}}

\newcommand{\ISV}{\mathbf{h}}
\newcommand{\NN}{\mathsf{N}\!\mathsf{N}}
\newcommand{\ANODE}{ANODE\xspace}
\newcommand{\NODE}{NODE\xspace}
\newcommand{\ISVNODE}{ISV-NODE\xspace}
\newcommand{\Inv}{\mathcal{I}}
\newcommand{\eq}{\,{=}\,}

\ifclean
\renewcommand{\COMMENT}[1]{{}}
\else
\usepackage{datetime}
\newcommand{\caution}{\the\day \ \monthname, \the\year \ \red{\bf Draft: do not distribute} }
\fi

\title{\bf A neural ordinary differential equation framework for modeling inelastic stress response via internal state variables}
\author{
R.E. Jones,\footnote{email: rjones@sandia.gov} \,
A.L. Frankel,
K.L. Johnson \\
{\it Sandia National Laboratories}
}
\ifclean
\date{}
\else
\date{\caution}
\fi

\begin{document}
\maketitle

\begin{abstract}
We propose a neural network framework to preclude the need to define or observe incompletely or inaccurately defined states of a material in order to describe its response.
The neural network design is based on the classical Coleman-Gurtin internal state variable theory.
In the proposed framework the states of the material are inferred from observable deformation and stress.
A neural network describes the flow of internal states and another represents the map from internal state and strain to stress.
We investigate tensor basis, component, and potential-based formulations of the stress model.
Violations of the second law of thermodynamics are  prevented by a constraint on the weights of the neural network.
We extend this framework to homogenization of materials with microstructure with a graph-based convolutional neural network that preprocesses the initial microstructure into salient features.
The modeling framework is tested on large datasets spanning inelastic material classes to demonstrate its general applicability.
\end{abstract}

\section{Introduction}

Relative to the certainty of the balance laws and the accuracy of the numerical methods, the fidelity of constitutive models is typically the weak point of predictive simulation for solids.
This is in part due to the fact that solid material behavior is complex and frequently influenced by microstructure.
Generally it can span a range of regimes with elastic, viscous, damage, and/or plastic characteristics, as well as have anisotropy and/or temperature-dependence.
This complexity often leads to models with significant discrepancies, which can be expensive to evaluate.

The response of many technologically relevant solid materials is inelastic, which means the material is dissipative and its stress and other aspects of its response depend on the deformation history.
This history-dependence is typically modeled with (a) hereditary integral formulations where kernels encode fading memory \cite{coleman1961foundations,lubliner1969fading}, or (b) with differential models with evolving state variables \cite{coleman1967thermodynamics,kratochvil1969thermodynamics}.
Examples of the first type, which result in Volterra integral equations, are the non-linear relaxation kernel of traditional viscoelasticity models \cite{coleman1961foundations,lubliner1969fading} and the memory kernel of Mori-Zwanzig  formalism employed primarily in fluid mechanics \cite{chorin2000optimal,li2014construction,parish2017dynamic} and particle systems \cite{adelman1976generalized,wagner2003coupling}.
The second route, where the present state of the material is expanded beyond observable deformation \cite{coleman1967thermodynamics}, is arguably more extensible since it has been applied to a much wider class of solid material response than viscoelasticity \cite{kratochvil1969thermodynamics,mcdowell2005internal,horstemeyer2010historical}.
In this second category the observable state is augmented with additional state variables that evolve via ordinary differential equations (ODEs).
This is in contrast to hypoelasticity \cite{truesdell2004non} where stress, itself, is governed by an ODE.
Although oftentimes physically-motivated, such as molecular vibrations, order parameters, or dislocation densities, these state variables are typically hidden from direct observation and may be incomplete or inappropriate to accurately predict the observable behavior.
Generally these type of models are referred to as \emph{internal state variable} (ISV) models and follow from Coleman and Gurtin's seminal work \cite{coleman1967thermodynamics}.
Coleman and Gurtin acknowledged that the delineation between observable and hidden can be arbitrary but gave the perspective that the hidden internal variables do not have to be ultimately observable; they must merely represent the process of interest phenomenologically.
In fact, Kr\"{o}ner \cite{kroner1965how} stated that a complete description of microstructure arrangement is unnecessary as long as the macroscale representation is complete in the sense that it is predictive.
This fundamental concept is at the core of model reduction and homogenization where macroscale processes depend on summary statistics of microscopic states.
The proposed inferred state variable neural ODE (\ISVNODE) model is motivated by this perspective.

\subsection{Internal state variable theory}
The applications of the ISV constitutive framework to solid material behavior are myriad.
Closely following the publication of the Coleman-Gurtin ISV theory \cite{coleman1967thermodynamics},  Kratochvil and Dillon \cite{kratochvil1969thermodynamics}  and Rice \cite{rice1971inelastic} framed elastoplasticity in the ISV framework, Bhandari and Oden \cite{bhandari1973unified} provided a viscoplastic ISV theory, and Perzyna developed an ISV model of ductile failure \cite{perzyna1986internal} based on evolution of porosity (what is now known as a {\it damage} model).
Later Reese and Govindjee \cite{reese1998theory} provided a particularly illuminating example of the application of ISV theory.
They translated and generalized the standard rheological model for viscoelasticity to finite deformations using an ISV plasticity framework based on the work of Simo \etal in finite deformation associative elastoplasticity \cite{simo1992associative} and viscoplasticity \cite{simo1992algorithms}.
Reese and Govindjee employed the now ubiquitous multiplicative decomposition of deformation gradient and an additive decomposition of the free energy potential, and demonstrated the formulation with oscillatory loading shear and  steady loading creep tests.

The breadth and depth of ISV constitutive modeling developments has motivated a number of reviews of the field.
In 1983, Germain \cite{germain1983continuum} published a summary and survey ISV elastoviscoplasticity theory.
Later, in 2005, McDowell \cite{mcdowell2005internal} published an in-depth and insightful review of the ISV framework for incorporating experimental data of irreversible/path-dependent behavior and gave an example of metal viscoplasticity.
That work provided a detailed discussion of the multiscale aspects of the theory and physically-motivated mesocale mechanics in general, including the assumption of locality for a representative volume element (RVE) in terms of the correlation length of relevant microscopic fields and other statistical aspects.
In 2010, Horstemeyer and Bammann \cite{horstemeyer2010historical} wrote a historical survey the development of ISV theory tying it to the inception of irreversible thermodynamics by Onsager \cite{onsager1931reciprocalI,onsager1931reciprocalII} and later Eckart \cite{eckart1940thermodynamics,eckart1948thermodynamics}.
Horstemeyer and Bammann also made a comprehensive review of the applications of ISV in solid mechanics which ranged across plasticity, viscosity/creep, and damage as well as multiphase, composite, biological and particulate materials.

The ISV theory is clearly an effective framework but still suffers from the primary difficulty of traditional constitutive modeling: discrepancies and model-form errors due to calibrating preconceived functional forms and assumed state variables.
In fact, there is no generally applicable prescription for determining appropriate/representative variables that quantify internal states \cite{mcdowell2005internal}.
ISV approaches oftentimes result in overly-complicated models that are costly to evaluate in large-scale simulation.
In contrast, machine learning models, like other general statistical models, are largely data-driven and applicable through a general and extensible methodology.
Furthermore current efforts \cite{ling2016machine,raissi2019physics,lee2019deep,wang2020incorporating,karniadakis2021physics,linka2021constitutive,masi2021thermodynamics} to hybridize machine learning models for physics with the principles used to develop traditional models are enabling the best of both approaches: efficient, expressive data-driven and traditional physics-informed modeling.

\subsection{Neural networks}
There have been considerable developments using neural networks (NN) \cite{tsoi1997discrete,yu2019review} for modeling dynamical systems and related tasks, such as sequence learning for natural language parsing \cite{lipton2015critical} and  normalizing flows used for generative modeling \cite{kobyzev2020normalizing}.
Most of these developments fall in the category of recurrent NNs (RNNs), such as long short-term memory (LSTM) \cite{hochreiter1997long} and the gated recurrent unit (GRU) \cite{cho2014learning}, which have memory to capture the causality of most time signals.
Based on the connection between the ResNet \cite{he2016deep}, which predict differences between states, and traditional discretizations of ODEs, new architectures have been recently proposed.
In 2018, Chen \etal \cite{chen2018neural}  introduced the Neural Ordinary Differential Equation (NODE) where the right-hand-side driving term of a system of ODEs is represented by a general multi-layer perceptron (MLP),
This formulation incorporates the time step scaling of the dynamics, which is missing in RNNs.
Closely following this development, in 2019 Dupont \etal \cite{dupont2019augmented} produced a major improvement with the augmented neural ODE (\ANODE).
In an \ANODE additional degrees of freedom are introduced to untangle the flows of the observed data; this effectively alleviates the necessity for paths to cross in order to represent some datasets.
This is the main enhancement the original NODE since trajectory crossing is disallowed in non-autonomous ODEs by the Picard–Lindel\"{o}f theorem \cite{coddington1955theory,arnold1973ordinary}.
This expansion of the state space of the dynamical system enabled learning complex dynamics with simpler flows, apparently generalized better, achieved lower losses with fewer parameters, and was more stable in training.
In 2020 Rackauckas \etal \cite{rackauckas2020universal} developed a similar framework, called Universal Differential Equations (UDEs), based on universal approximation properties and focused on physical applications.
In fact, NODEs and the like have been shown to have a universal approximation property \cite{rubel1981universal,lu2019deeponet,teshima2020universal}, like the related MLPs \cite{hornik1989multilayer,scarselli1998universal}.
They also have a Bayesian extension \cite{dandekar2020bayesian} that embeds uncertainty in the model predictions.
The ODE-based NN architectures are adaptable, generalizable, and have the distinct appeal of resembling classical methods of representing and simulating time-continuous dynamical systems.

There have been a number of developments specific to applying NN to modeling evolving physical systems.
Xiu and co-workers have been particularly active in this area.
Fu \etal \cite{fu2020learning} created a generalized Langevin model resulting from a discrete and finite Mori-Zwanzig memory kernel with a ResNet architecture; and, Qin \etal \cite{qin2021data} reframed the time input of non-autonomous differential equations into that of a time-parameterized general forcing input to model a general class of evolutionary behavior.
A number of other developments have focused on control applications.
Drgona \etal \cite{drgona2020spectral} developed a RNN type model of buildings subject to real world heating requirements for model-based  predictive feedback control and compared to alternative strategies.
Drgona and co-workers \cite{drgona2020spectral} also made a spectral analysis of deep networks with common activation functions and categorized their dynamic stability.

Physics constrained approaches have received considerable attention and development with contributions such as Physics Informed Neural Networks \cite{lagaris1998artificial,raissi2019physics}.
Specific to material modeling, the Tensor-Basis Neural Network (TBNN) \cite{ling2016machine,jones2018machine,frankel2020tensor,peters2020s} was developed to embed the symmetries imposed by material frame indifference (equivariance).
As an adaptation of classical representation theory to neural network architecture, in a TBNN stress, or any other tensorial output, is represented by a finite sum of trainable NN coefficient functions of the scalar invariants of the inputs paired with the tensor basis elements of the inputs.
Two approaches have been introduced to construct a TBNN: (a) an implicit TBNN, which is trained directly to the stress output and the coefficients are inferred   \cite{ling2016machine}, and (b) an explicit TBNN, where the coefficients are solved from the stress and strain invariants then the network is trained to these coefficients  \cite{frankel2020tensor}.
The TBNN approach where symmetry is embedded in the formulation, as opposed to learned, has been shown to reduce the training burden \cite{ling2016machine}.

A few notable developments are particular to inelastic solid behavior.
Xu \etal \cite{xu2020inverse} developed a NN model of viscoelastic behavior and trained it via optimization constrained by solutions to the boundary value problem of the application.
Their model resembles hypoelasticity formulated by stress increment as a function of current stress and strain and was demonstrated on simplified geomechanics problems.
Masi \etal \cite{masi2021thermodynamics} also developed a hypoelastic NN representation based on modeling increments in stress, a dissipation function, and traditionally defined internal state variables.
Recently, Logarzo \etal \cite{logarzo2021smart} used an RNN to model homogenized inelastic response of an elastic-plastic matrix with a single stiff elastic inclusion.
In contrast, Vlassis and Sun \cite{vlassis2020sobolev} embedded a key feature of traditional plasticity theory with their implementation of a  NN yield function.
They demonstrated superior performance to alternative RNN formulations that lack an explicit yield surface with their model.
Also relevant to this work, Teichert \etal \cite{teichert2019machine} demonstrated that thermodynamic potentials can be inferred by an NN trained on derivative data, such as stress.

\subsection{An internal state motivated neural network}
The goal of this work is to create a representation suitable to modeling a wide class of dissipative materials that obeys physical principles, such as frame invariance and the second law of thermodynamics; and that treats hidden states, like those related to damage, in a flexible and data-driven manner.
In this representation the state of the material is inferred from data, not defined by preconceived quantities, such as plastic-strain and damage, that require unambiguous definition and measurement.
As in the ISV framework and our previous work \cite{jones2018machine}, the proposed model has interpretable components that are analogs to: (a) the traditional flow rule, which evolve internal state based on strain loading, and (b) a stress model, mapping state to observable stress.
The evolution of the inferred state vector is handled with traditional time-integrators compatible with current large-scale simulators.
Furthermore, the state-space can be built with complexity appropriate for the available data and physical process, as with the \ANODE.
We expand the treatment of homogeneous materials, where the state vector is completely inferred, to those with microstructure where the relevant structural features are discovered with a hybrid convolutional neural network (CNN)-RNN architecture devised in previous work  \cite{frankel2019predicting,frankel2020prediction,vlassis2020geometric,frankel2021mesh}.
In contrast to previous developments, in this work we couple a graph-based CNN (GCNN) applied directly to the unstructured mesh data \cite{frankel2021mesh} to the \ISVNODE\ dynamical model.
The GCNN processes the initial microstructure to a latent set of initial features that augment the state vector in our  \ANODE-like ISV framework.

In the following sections we develop and demonstrate the proposed \ISVNODE framework.
In \sref{sec:theory} we outline Coleman-Gurtin internal state variable theory and connect it to Augmented Neural ODEs in \sref{sec:architecture} to put the proposed architecture in context.
We apply the \ISVNODE framework to modeling viscoelastic and elastoplastic homogeneous materials and similar materials with microstructure.
The training methodology is described in \sref{sec:training}, and the multitude of training data is discussed in \sref{sec:data}.
With these in hand we explore the qualities of the \ISVNODE and  demonstrate its performance in \sref{sec:results}.
We conclude with a discussion of the developments, open questions, and future work in \sref{sec:conclusion}.

\section{Theory} \label{sec:theory}

Our goal is to create a constitutive modeling framework for the stress response of inelastic materials based solely on thermodynamic principles and  observable, measurable quantities, namely: (a) the material motion $\xb = \chib(\Xb,t)$, from reference position $\Xb$ to current position $\xb$ as a function of time, and (b) the Cauchy stress $\Tb$.
The classical Coleman and Gurtin \cite{coleman1967thermodynamics} internal state variable (ISV) theory postulates that general inelastic response can be described with a few canonical response functions which are dependent on the current state of a deformation measure, temperature, and a collection of additional, internal state variables.
The Helmholtz free energy $\Psi$ determines stress and entropy through its partial derivatives with respect to deformation and temperature, while the heat flux and evolution of the ISVs need to be specified independently from the free energy.

In this work, we restrict the developments to isothermal processes.
In this case, the Coleman-Gurtin ISV framework reduces to a stress response
\begin{equation} \label{eq:stress_rule}
\Sb = \partialb_\Eb \hat{\Psi}|_\ISV = \hat{\Sb}(\Eb, \ISV)
\end{equation}
and a flow of the internal state variables
\begin{equation} \label{eq:flow_rule}
\dot{\ISV} = \hat{\fb}(\Eb, \ISV)
\end{equation}
which are functions of observable strain $\Eb$ and the additional internal state variables $\ISV$.
Many traditional models of inelastic solids \cite{truesdell2004non,simo2006computational,gurtin2010mechanics,silhavy2013mechanics} fit into this framework.
Here we have chosen the second Piola-Kirchhoff stress $\Sb$ as the stress measure, which is related to the Cauchy stress $\Tb$ by $ \det(\Fb) \Tb = \Fb \Sb \Fb^T $, and the Lagrange strain, $ \Eb = \frac{1}{2} \left( \Fb^T \Fb - \Ib \right)$, where $\Fb = \partialb_\Xb \chib(\Xb,t)$ is the deformation gradient.
These stress and strain measures are energetic duals and inherently invariant to superposed rigid motion or changes of material coordinate frame.
Both arguments, $\Eb$ and $\hb$, are present in both response functions by Truesdell's principle of equipresence \cite{truesdell1959rational,truesdell1960classical}, which can be considered as a design principle for constitutive models.
Truesdell and Toupin \cite{truesdell1960classical} postulate a number of general principles for constitutive models including dimensional independence, spatial and material invariances and other symmetries which we also follow.

The second law of thermodynamics is particularly relevant to inelastic materials since it constrains allowable stress responses.
The Clausius-Duhem inequality \cite{silhavy2013mechanics} reduces to the rate of change of the free energy being bounded by the stress power
\begin{equation}
\dot{\Psi} \le \Sb \cdot \dot{\Eb}
\end{equation}
in the isothermal case.
This principle, together with the definition of stress $\Sb$, \eref{eq:stress_rule}, and the energy balance for isothermal process
\begin{equation}
\dot{\Psi} = \Sb \cdot \dot{\Eb} + \partialb_\hb \Psi \cdot \fb \ ,
\end{equation}
constrain the flow of the internal state variables
\begin{equation} \label{eq:isv_inequality}
\partialb_\hb \Psi \cdot \fb \le 0 \ .
\end{equation}
This result, and the definition of stress $\Sb$ as the partial derivative of the free energy $\Psi$ with respect to the strain $\Eb$ with the internal variables $\hb$ {\it fixed}, implies the ISVs characterize inelastic behavior and, in particular, the change of state associated with irreversible deformation.

Other than \eref{eq:isv_inequality}, the Coleman-Gurtin theory does not provide general prescriptions for the form of the function $\fb$ driving the evolution of the ISVs.
We augment the arguments of $\fb$ with the strain rate $\dot{\Eb}$:
\begin{equation}
\dot{\ISV} = \hat{\fb}(\Eb,\dot{\Eb},\ISV) \ ,
\end{equation}
so that the rate dependence does not need to be derived from the history $\Eb(t)$ and allow the flow rule to have an explicit sense of loading direction.
We chose to have state variables $\ISV$ to be nominally in the reference configuration, so the material time derivative is an appropriate rate.
Since many other objective rates \cite{haupt1989application} are available, this selection is effectively a model form choice, like our choice of stress and strain measures; however, it is not a particularly restrictive one.
The set of internal state variables, $\ISV$, does not necessarily have a tensorial character \cite{mcdowell2005internal}.
Lacking general principles motivating more complex choices, we assume the hidden state $\ISV$ can be characterized by a (non-tensorial) collection of scalars in the reference configuration.
This implies that the state evolution $\dot{\ISV}$ is driven by the invariants, $\hb$, and the invariants of the strain loading, $\Inv(\Eb,\dot{\Eb})$:
\begin{equation} \label{eq:isv_flow}
\dot{\ISV} = \hat{\fb}(\Inv(\Eb,\dot{\Eb}),\ISV)
\end{equation}
Initial conditions for the hidden states $\ISV(t\eq 0)$ can be set to $\mathbf{0}$ lacking other information.
If additional data is available, for example, a fixed state such as temperature or an initial microstructure, the state $\hb$  can be augmented with this information as we will show in the next section.

Stress model $\hat{\Sb}(\Eb,\hb)$ has a number of valid formulations.
A potential-based model $\hat{\Psi}$ follows directly from the Coleman-Gurtin theory
\begin{equation} \label{eq:pot_stress}
\Sb = \partial_\Eb \hat{\Psi}(\Eb,\hb)
\end{equation}
This requires inferring the potential $\Psi$ from the observable stress and deformation data.
A tensor-basis formulation
\begin{equation} \label{eq:tb_stress}
\Sb = \sum_i \hat{\sigma}_i(\Inv(\Eb,\hb)) \Bb_i
\end{equation}
is an alternative that has the possible advantages of being a smoother model and having simpler dependence on the strain since the basis $\Bb_i$ embeds some of the functional dependence.
In this form, $\hat{\sigma}_i$ are functions of the invariants of $\Eb$ and $\hb$ and the basis $\Bb_i$ is constructed from $\Eb$ (and possibly $\dot{\Eb}$, see \aref{app:tb} for details).
Without access to the potential this formulation  cannot enforce the Clausius-Duhem inequality \eref{eq:isv_inequality} directly.
This will be discussed further in the next section.
A component-based representation
\begin{equation} \label{eq:comp_stress}
\Sb = \sum_{(ij)} \hat{s}_{(ij)} (\Eb,\hb) \eb_{(ij)}
\end{equation}
presents a second alternative, where $\eb_{(ij)} \equiv \frac{1}{2}( \eb_i \otimes \eb_j + \eb_j \otimes \eb_i)$ are the symmetric Cartesian basis dyads in the reference configuration.
This formulation is similar to the tensor-basis formulation; however, the basis is fixed and the components $\hat{s}_{(ij)}$ are function of the components of strain, not its invariants.

\section{Architecture} \label{sec:architecture}

As discussed in the preceding sections, the primary shortcoming of the general ISV modeling framework is: it not clear what the internal state variables $\hb$ should be nor even what dimensionality of this set should be.
To alleviate this shortcoming we developed a data-driven Neural Network (NN) analog to the ISV framework based on an enhanced Neural ODE (NODE).

\subsection{The \ANODE} \label{sec:anode}
To put our developments in context we first outline the basic form of the Augmented Neural ODE (\ANODE) \cite{dupont2019augmented}.
Briefly, its objective is to train a model of a time-dependent process $\yb = \yb(t)$ assuming its true form is given by an ODE:
\begin{equation} \label{eq:ode}
\dot{\yb} = \fb(\yb,t) \ .
\end{equation}
Training is based on observations of $\yb$ over time $t$ and $\fb(\yb,t)$ is modeled with a multilayer perceptron (MLP), a deep, densely-connected feed-forward NN,  \cite{rosenblatt1961principles,goodfellow2016deep}.
Each layer of the MLP translates its inputs $\xb$ to outputs with a parameterized affine transformation followed by a non-linear transformation:
\begin{equation}
\xb_{i+1}(t) = a(\Wb_i \xb_i(t) + \bb_i) \ ,
\end{equation}
where  $\Wb_i$ is the  trainable weight matrix of the $i$-th layer, $\bb_i$ is the trainable bias, and $a$ is the preselected non-linear activation function applied element-wise.
The input $\xb_0(t) = [ \yb(t), t, \hb(t) ]^T$ to the NN approximation of the dynamics, $\NN(\xb_0)$, is formed from the observable state vector $\yb$ augmented with time $t$ and  additional hidden variables $\hb$.
This augmentation, not found in the original Neural ODE \cite{chen2018neural}, allows trajectories to be simpler, in a sense, and not to cross each other or themselves by expanding the dimensionality of the output space.
The dimension of $\hb$ determines the size of latent space needed to represent the process $\yb(t)$ and, in part, the complexity of the approximation.
The other primary contributor to the complexity of the representation is the number of layers $n$ in the MLP.
The output of the final layer of $\NN(\xb_0(t))$ is $\xb_n = [ \fb, 1, \dot{\hb} ]$.
Typically the application of the non-linearity $a$ is omitted for this final layer to allow for linear mixing of the product of previous layers.
This function can then be integrated with a standard numerical integrator, for instance with an explicit midpoint rule
\begin{equation}
\xb(t+\delt) = \xb(t) +  \delt \, \NN(\xb(t+1/2\delt),t+1/2 \delt)
\end{equation}
where $\xb(t+1/2\delt) = \xb(t) +  1/2 \delt \, \NN(t)$.
As opposed to a RNN, in this architecture the scaling with time is explicitly accounted for by the embedded time integrator.
In general, where no additional information is available about the initial state, the initial conditions for $\hb(t\eq 0)$ are set to zero.
Training is done with standard back-propagation \cite{rumelhart1986learning,robbins1951stochastic} or via adjoint methods \cite{chen2018neural}.
Note that both the original NODE formulation \cite{chen2018neural} and the Augmented NODE \cite{chen2018neural} in effect augment the apparent state vector $\yb$ with time $t$ so that the MLP approximates both the dependence of $\fb$ on $\yb$ and $t$ with time invariant weights $\{ \Wb_i \}$ and biases $\{ \bb_i \}$.

\subsection{The \ISVNODE} \label{sec:isvnode}
\begin{figure}
\centering
\subfloat[potential-based]
{\includegraphics[width=0.70\textwidth]{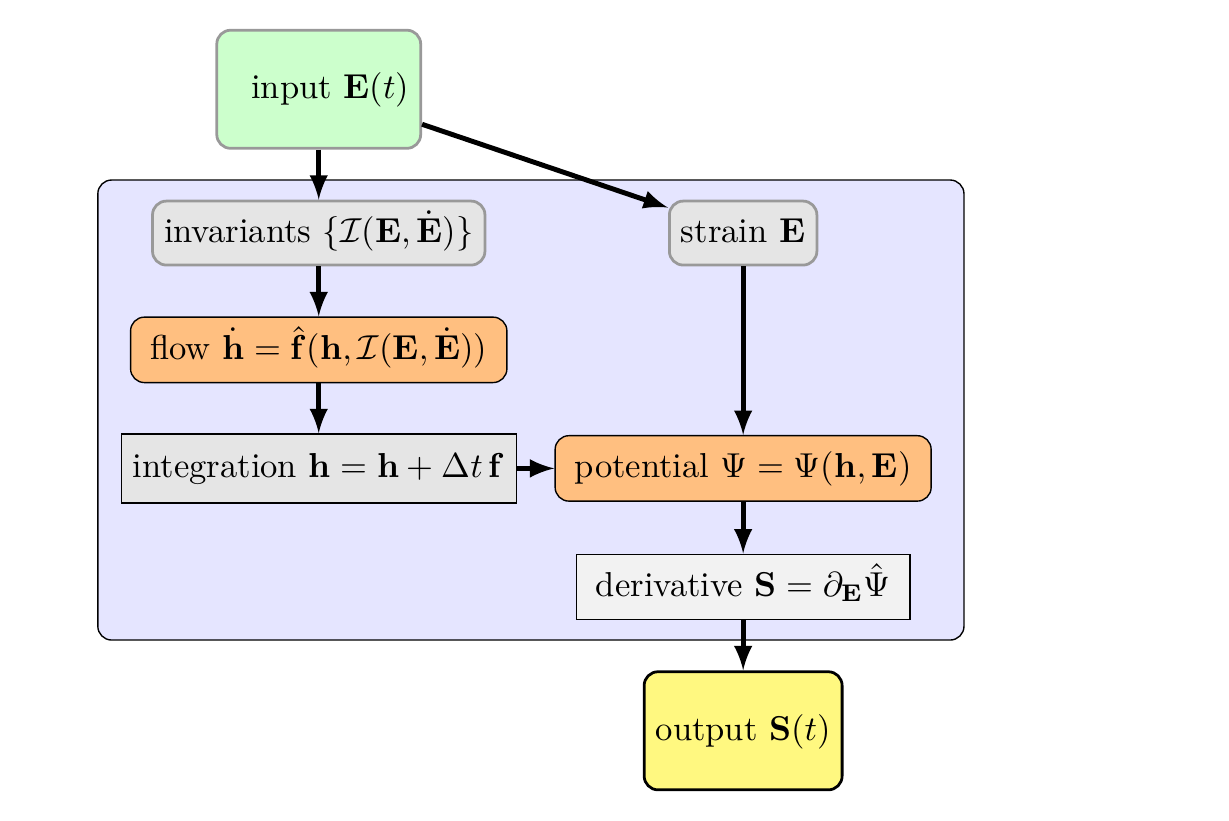}}

\subfloat[tensor basis]
{\includegraphics[width=0.70\textwidth]{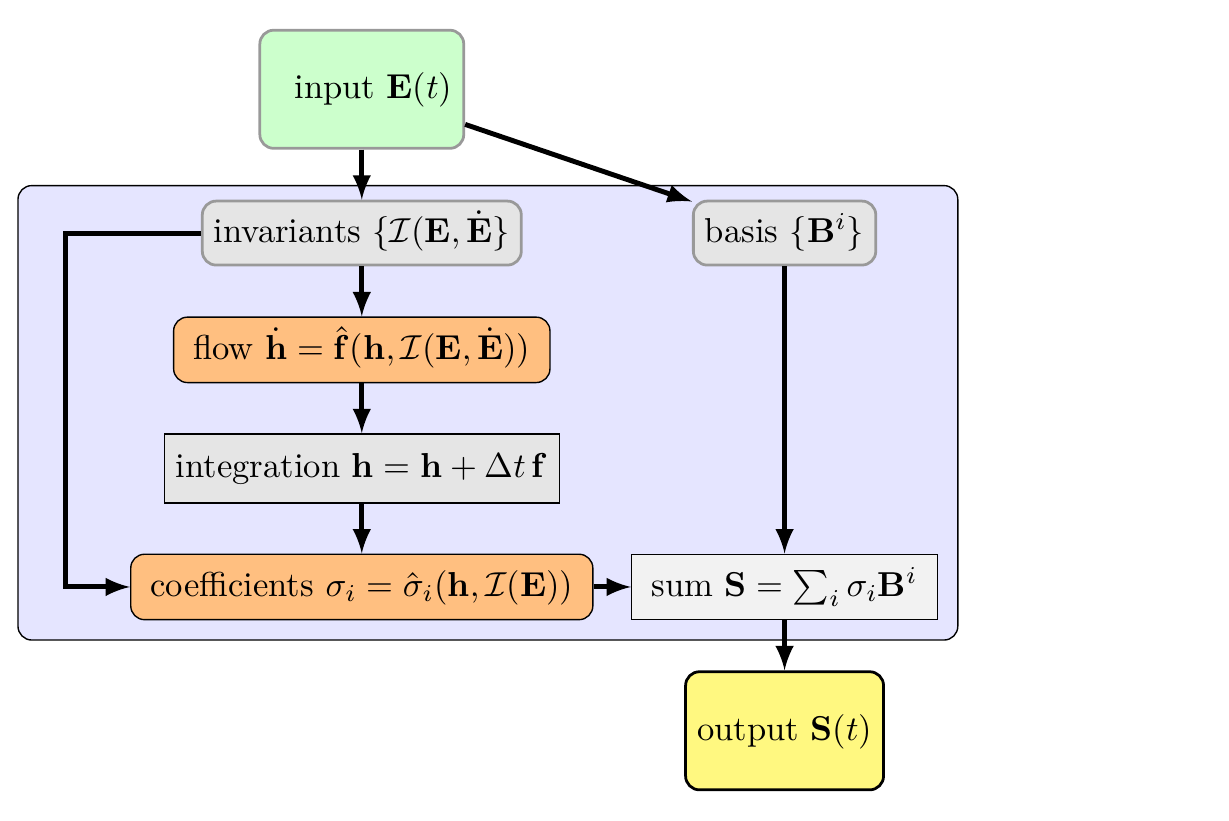}}

\caption{\ISVNODE\ architecture:
(a) potential-based stress,
(b) tensor basis stress formulations.
Colors denote:
green: input,
orange: trainable (state evolution and state-to-output) MLPs,
gray: non-trainable operations, and
yellow: output.
Note the size of $\ISV$ is user selected and  $\ISV(t\eq 0) = \mathbf{0}$ initial conditions are given to the integrator.
}
\label{fig:homogeneous_architecture}
\end{figure}

Clearly the flow rule \eref{eq:isv_flow} resembles the non-autonomous ODE, \eref{eq:ode}, the \ANODE\ is built on; however, we need a number of enhancements to  adapt this methodology to the ISV framework.
With reference to the schematic in \fref{fig:homogeneous_architecture}, first we replace explicit dependence on time $t$ in $\fb$ in \eref{eq:ode} with dependence on the strain loading.
Specifically we use the invariants of the strain and strain rate $\Inv(\Eb,\dot{\Eb})$, which can be found in \aref{app:tb}, as inputs to form a representation of $\hat{\fb}$ from \eref{eq:flow_rule}.
In addition to following the ISV framework and the Truesdell-Toupin design principles, this modification enables $\fb$ to readily generalize to arbitrary loading paths.
Second, we add an additional MLP to transform the output of the integration of the flow rule, \ie the hidden state $\hb$, and the strain $\Eb$ to the observable stress $\Sb$ following \eref{eq:stress_rule}.
The inclusion of strain $\Eb$ in the arguments of $\hat{\Sb}$ promotes $\dot{\hb} = \mathbf{0}$ when no dissipation is occurring.
We will also introduce a penalization of the related constraint, \eref{eq:isv_inequality}, to the loss function, discussed in the next section.

The potential-based \eref{eq:pot_stress}, tensor basis \eref{eq:tb_stress}, and component-based \eref{eq:comp_stress} variants each lead to slightly different architectures.
In the potential-based variant, illustrated in \fref{fig:homogeneous_architecture}a, the potential $\hat{\Psi}$ is represented by the second MLP, and its derivative with respect to strain produces the model stress $\hat{\Sb}$.
Note the potential is explicit in this formulation but never compared directly to data.
In the tensor basis variant, illustrated in \fref{fig:homogeneous_architecture}b, the basis coefficient functions $\hat{\sigma}_i (\hb, \Inv(\Eb))$ are represented with by a MLP and the outputs are summed with their basis elements to produce the model stress.
The component-based variant resembles the tensor basis variant in that the coefficient functions $\hat{S}_{(ij)}$  are approximated and summed with a fixed basis to produce the model stress, as in \eref{eq:comp_stress}.

The \ISVNODE architecture has a number of hyper-parameters:
$N_\fb$, the number of layers in the state-evolution MLP $\NN_\fb$;
$N_\Sb$, the number of layers in the stress MLP $\NN_\Sb$; and
$N_\ISV$, the number of hidden ISVs.
The choice of the number of hidden states $N_\ISV$ is of primary importance and determines, together with the sizes of $\Inv(\Eb,\dot{\Eb})$,  $\Inv(\Eb)$, and $\Sb$, the width of the MLPs.
Following the stability analysis of Drgona \etal \cite{drgona2020spectral} and some preliminary studies, we selected the $C^1$-smooth exponential linear unit (ELU) as the activation $a$ for both MLPs.

\begin{figure}
\centering
\subfloat[potential-based ISV-NODE with GCNN input]
{\includegraphics[width=0.72\textwidth]{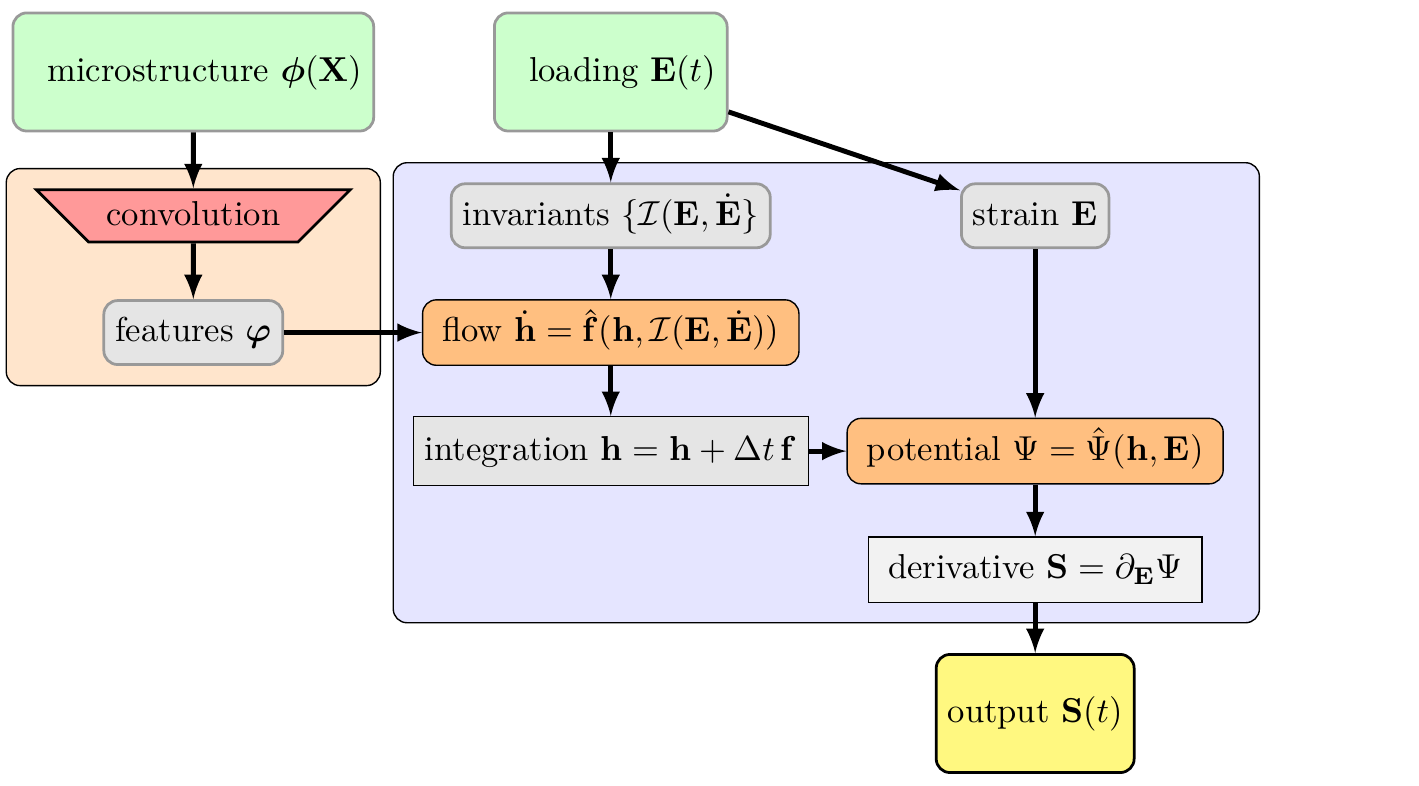}}
\subfloat[convolutional unit]
{\includegraphics[width=0.24\textwidth]{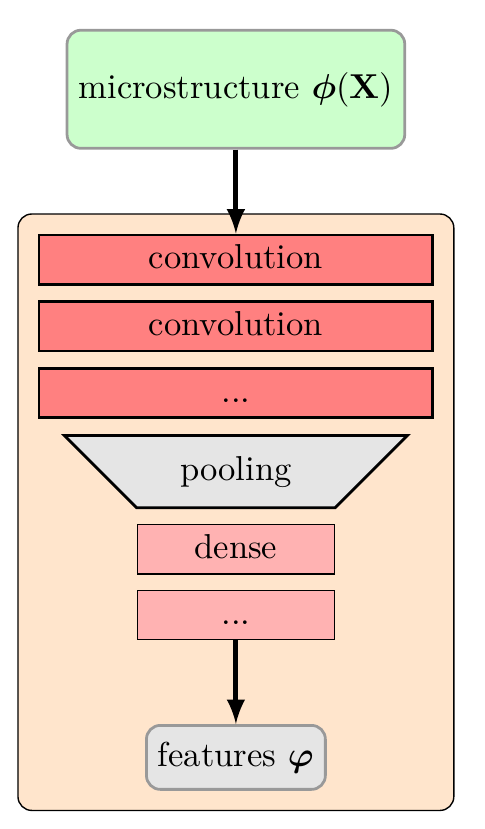}}
\caption{\ISVNODE\ architecture with microstructure input $\phi(\Xb)$: (a) overall architecture for a potential-based formulation, and (b) details of the convolutional unit.
Colors denote:
green: input,
red: convolution,
orange: trainable (state evolution and state-to-output) MLPs,
gray: non-trainable operations, and
yellow: output.
The convolution unit is circumscribed by an orange background and the NODE unit is circumscribed by a blue background.
Note the output of the convolutional unit, features $\varphib$, is concatenated with the initial conditions of the user selected state variables $\ISV$.
}
\label{fig:heterogeneous_architecture}
\end{figure}

As in the \ANODE we use $\ISV(t\eq 0)=\mathbf{0}$ initial conditions for the hidden state variables in the absence of data-derived state information.
We will demonstrate this is effective in modeling homogeneous materials.
For heterogeneous materials we augment the hidden state $\ISV$.
Following \crefs{frankel2019predicting,frankel2021mesh} and with reference to \fref{fig:heterogeneous_architecture}a, we reduce initial microstructural fields with a convolutional neural network (CNN).
We augment the material state, $\ISV$, with these microstructural features.
The CNN component that reduces the initial microstructure to structural features is trained simultaneously with the \NODE component.
These inferred features take the form of additional initial conditions for $\ISV$, unlike in our previous RNN-based formulations where this information was a constant input to the RNN.
As in \cref{frankel2021mesh} we employ a graph-based convolutional unit applied directly to the data on the unstructured discretization of the microstructure.

\fref{fig:heterogeneous_architecture}b illustrates the internal structure of the convolutional unit that processes the fields describing the initial microstructure $\phib(\Xb)$ into relevant features $\varphib$.
It consists of a number of graph convolutions applied to $\phib$, each operates on nearest neighbors elements as defined by the unstructured mesh used to compute the response.
Each convolutional layer is endowed with $N_\text{filters}$ independent filters.
After $N_\text{convolution}$ convolutions, global pooling is applied to reduce the output to dimension $N_\text{filters}$.
This output is fed in $N_\text{dense}$ densely connected layers and ultimately a final linear mixing layer, as in a MLP.
For this version of the convolutional unit the size of the structural features is equal to the number of filters $N_\text{filters}$.
\cref{frankel2021mesh} gives more details of GCNN architecture.

The \ISVNODE was implemented with TensorFlow \cite{tensorflow} and the GCNN with Spektral \cite{spektral}.

\section{Training} \label{sec:training}
As can be seen in \fref{fig:homogeneous_architecture} the inputs to the fundamental \ISVNODE are the strain $\Eb(t)$ and its rate $\dot{\Eb}(t)$ corresponding to the loading of the sample.
These are collected from the simulations used as a data source at discrete times $\{t_i = i \delt, i=0,N_\text{steps}\}$  where $\delt$ is fixed per sample but varies across the ensemble of data.
A fixed time-step per trajectory was chosen as a convenience for collecting data from the data generating simulator, it is not a restriction on the \ANODE.
For the state evolution NN this data is preprocessed into the corresponding joint invariants $\Inv(\Eb(t_i),\dot{\Eb}(t_i))$.
The output stress $\Sb(t_i)$ to be compared with the model output $\hat{\Sb}(t_i)$ in the loss function is obtained from the Cauchy stress of the simulations.
In the case of the heterogeneous simulations the output stress is the volume-averaged Cauchy stress.

To facilitate training the data is rescaled to be $\Oc(1)$.
The ensemble of inputs $\{ ( \Eb(t_i), \dot{\Eb}(t_i) )_k, \ k=1,N_\text{samples} \}$ and outputs $\{( \Sb(t_i) )_k, \ k=1,N_\text{samples}\}$ are rescaled by their respective maximums over the ensemble, \eg $\Sb \to \frac{1}{s_\Sb} \Sb$  where $s_\Sb = \max \left| ( S_{ij} )_k  (t_n) \right|$.
Since the data is sufficiently centered no shift of the mean was done.
The corresponding invariants are rescaled by powers of $s_{\tr{\Eb}} = \max \left| \tr\Eb \right|$ and $s_{\tr{\dot\Eb}} = \max \left| \tr\dot{\Eb} \right|$, \eg $\tr (\Eb^2 \dot{\Eb}) \to \frac{1}{s^2_{\tr{\Eb}} s_{\tr{\dot{\Eb}}} } \tr (\Eb^2 \dot{\Eb})$, for consistency.
The $N_\text{samples}$ samples were randomly split into training, testing and validation tranches in a 70/10/20 proportion.

We employed a standard mean square error (MSE) loss function
\begin{equation}\label{eq:loss}
L = \frac{1}{N_\text{samples}} \sum_{i=1}^{N_\text{samples}} \sum_{j=1}^{N_t}  \| \Sb_i(t_j) - \hat{\Sb}(\Eb_i(t_j),\dot{\Eb}(t_j)) \|^2
\end{equation}
applied to the stress data $\Sb$ and model response $\hat{\Sb}$.
(The root mean squared error (RMSE) is $\sqrt{L}$.)
To enforce the dissipation constraint imposed by the second law, we augment the loss $L$ with a penalty
\begin{equation}\label{eq:penalty_loss}
L_\varepsilon = L + \varepsilon g
\end{equation}
where $\varepsilon$ is the penalty hyperparameter and $g$ is the constraint function.
The form of constraint is determined by the stress model.
For the potential-based formulation of the stress, the constraint takes the form
\begin{equation}\label{eq:pot_penalty}
g = R(\partialb_\hb \Psi \cdot \fb)
\end{equation}
based on inequality, \eref{eq:isv_inequality}, where $R$ is the Macauley bracket, which is identical in form to the ReLU activation function, $R(x) = x$ if $x > 0$ and $R(x) = 0$ otherwise.
In the tensor basis and component-based stress formulations, the potential is not available so we chose to match the expended power
\begin{equation}\label{eq:tb_penalty}
g = \| (\Sb - \hat{\Sb})\cdot \dot{\Eb} \|^2
\end{equation}
as the constraint.
Note with a tensor-basis formulation, \eref{eq:tb_stress} allows \eref{eq:tb_penalty} to be written in terms of the coefficient functions $\hat{\sigma}_i$.
Furthermore,  \cref{jones2018machine} showed how to construct a strictly dissipative flow rule for assumed plastic strain ISV using selective powers of the input invariants; however, we have left adapting this to the generality of the \ISVNODE for future.

The training scheme to minimize the loss, $L_\epsilon$, involved a standard stochastic gradient descent algorithm, Adam \cite{kingma2014adam}, with an initial learning rate of 0.001 and batch size 64 for the homogeneous data.
Given the size of the microstructure inputs, the models for the heterogeneous data were trained with a batch size of 1, as in \cref{frankel2021mesh}.
Sequential training was used to ensure convergence, where the model was trained to the initial steps of the evolution $N_\text{steps}^i$, then to incrementally more of the evolution till a target portion of the training evolution was used.
A typical schedule was $N_\text{steps}^i = \{ 40, 80, 120, 200 \}$ where
early stopping based on a target accuracy was employed.
The target accuracy was increased as more of the evolution was included in the training.
Note a parallel approach where time-batched training sets would be used independently \cite{gunther2020layer} was not feasible since the internal state $\hb$ evolves and we assume $\ISV=\mathbf{0}$ only at $t=0$.

\section{Data} \label{sec:data}

We use training data drawn from two general classes of inelastic material behavior: viscoelasticity and elastoplasticity.
As in other work, typical, well-proven models of this behavior serve as stand-ins for sufficient experimental data as data fusion and other techniques are developed.
In addition to creating response data for homogeneous systems undergoing homogeneous deformations, we generated response data for large scale representative volumes  with pores or inclusions.
For this data our goal was to predict the homogenized response based on the initial microstructure.
The Sierra simulation suite \cite{sierra} was used to generate all training and testing data.

\subsection{Viscoelastic material} \label{sec:upm}

For a traditional model of a common engineering polymeric material, we employed a Universal Polymer Model (UPM) \cite{adolf2009simplified} for Sylgard 184 silicone, a lightly cross-linked, flexible, isotropic elastomer \cite{long2017linear}.
The UPM is a viscoelastic model of the hereditary integral type:
\begin{align}
\Tb &=
(K - K_\infty) \Ib \int_0^t \left( f_K(t-s) \tr \dot{\epsilonb} (s) \right) \, \mathrm{d}s + K_\infty \Ib \tr \epsilonb \\
&+ 2 (G - G_\infty) \int_0^t    \left( f_G(t-s) \dev \dot{\epsilonb} (s) \right) \, \mathrm{d}s + 2 G_\infty \dev \epsilonb  \nonumber
\end{align}
based on a bulk($K$)/shear($G$) split.
The strain measure $\epsilonb$ is given by the integration of the unrotated rate of deformation $\Db = \frac{1}{2} ( \grad_\xb \vb  + \grad^T_\xb \vb )$
\begin{equation}
\epsilonb = \int_0^t \Rb^T(s) \Db(s) \Rb(s) \, \mathrm{d}s
\end{equation}
where $\Rb$ is the rotation tensor from the polar decomposition of the deformation gradient $\Fb$.
The relaxation kernels $f_K$ and $f_G$ are represented with Prony series with 20 relaxation times ranging from 1 $\mu$s to 3160 s.
The instantaneous bulk and shear moduli, $K=$ 920 MPa and $G=$ 0.362 MPa, and equilibrium bulk and shear moduli, $K_\infty=$ 920 MPa and $G_\infty$=0.084 MPa,  and all other parameters are given in \cref{long2017linear}.

\subsection{Elastoplastic material} \label{sec:J2}

Aluminum was chosen as a typical elastic-plastic material as represented by a J2 model \cite{lubliner2008plasticity}.
The stress $\Sb$ is given by a linear elastic rule:
\begin{equation} \label{eq:stress}
\Sb = \Cbb : \Eb_e
\end{equation}
where ``:'' is a double inner product that allows the 4th order elastic modulus tensor $\Cbb$ to map the elastic strain $\Eb_e$ to the stress $\Sb$.
The elastic logarithmic strain is derived from a multiplicative split of the deformation gradient $\Fb= \Fb_e \Fb_p$.
For an isotropic material, like common aluminum, the components of $\Cbb$ reduce to
\begin{equation}
[ \Cbb ]_{ijkl} = \frac{E}{(1+\nu)} \left( \frac{\nu}{(1-2\nu)} \delta_{ij}\delta_{kl} + \frac{1}{2} (\delta_{ik}\delta_{jl} + \delta_{il}\delta_{jk}) \right)
\end{equation}
which depend only on Young's modulus $E=$ 59.2 GPa and Poisson's ratio $\nu=$ 0.33.

The plastic flow is derived from the von Mises yield condition
\begin{equation}
\sigma_\text{vm}(\Sb) - \check{\sigma}(\epsilon_p) \le 0
\end{equation}
which limits the elastic regime to a convex region in stress space and offsets the elastic strain $\Eb_e$ from the total strain.
Here $\sigma_\text{vm} = \sqrt{\frac{3}{2} \sb\cdot\sb}$ is the von Mises stress where $\sb = \Sb - \tr(\Sb) \Ib$ is the deviatoric part of $\Sb$, and $\epsilon_p$ is the equivalent plastic strain, which is a measure of the accumulated plastic strain computed from the plastic velocity gradient $\Db_p$
\begin{equation}
\epsilon_p = \int_0^t \sqrt{\frac{2}{3} \Db_p(s) : \Db_p(s)} \, \mathrm{d}s \ .
\end{equation}
The yield limit $\check{\sigma}$ is given by a Voce hardening law
\begin{equation}
\check{\sigma} = Y + H ( 1 - \exp(-\alpha \epsilon_p) )
\end{equation}
with parameters: initial yield $Y=$ 200.0 MPa, hardening $H=$ 163.6 MPa, and saturation exponent $\alpha=$ 73.3.

\subsection{Homogeneous response} \label{sec:homo_response}

\begin{figure}
\centering
\subfloat[viscoelastic UPM]
{\includegraphics[width=0.45\textwidth]{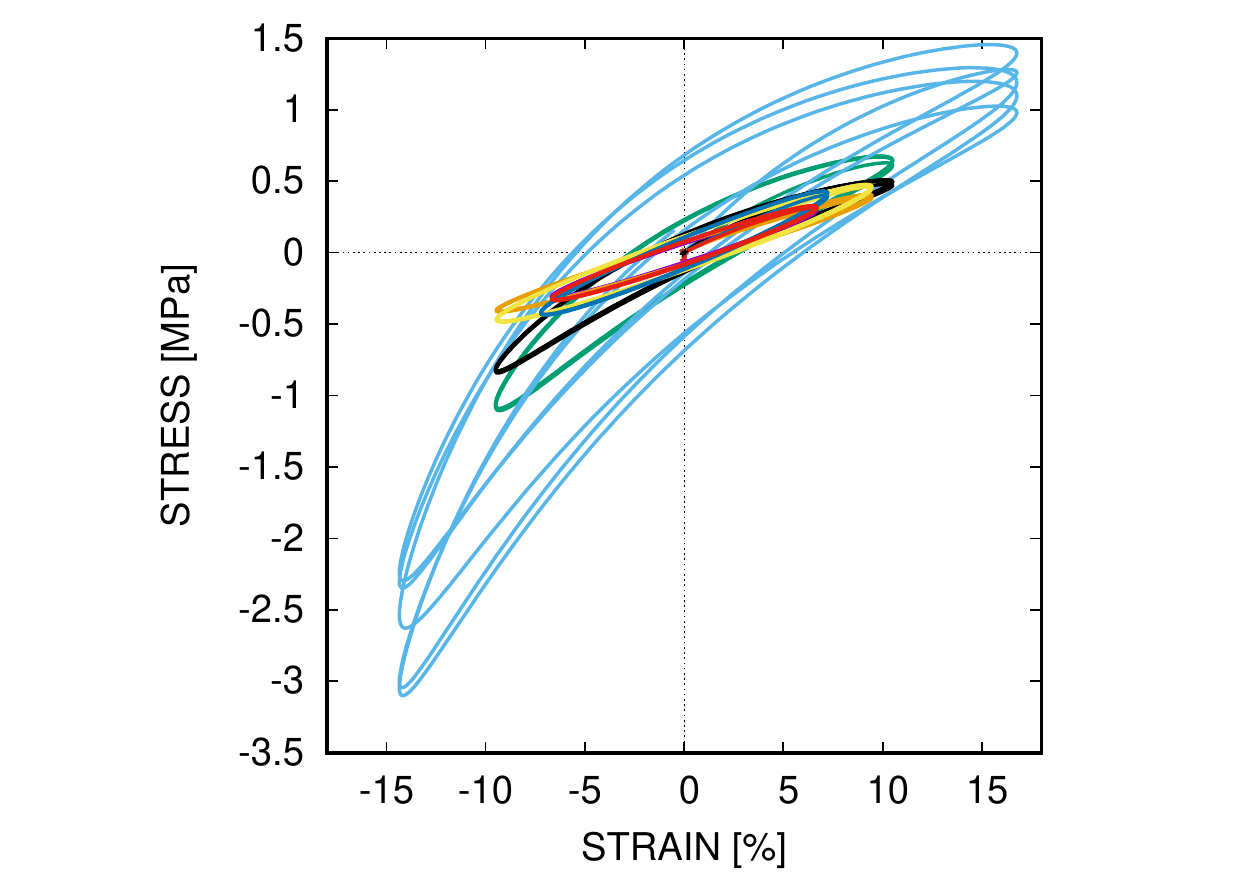}}
\subfloat[elastic-plastic J2]
{\includegraphics[width=0.45\textwidth]{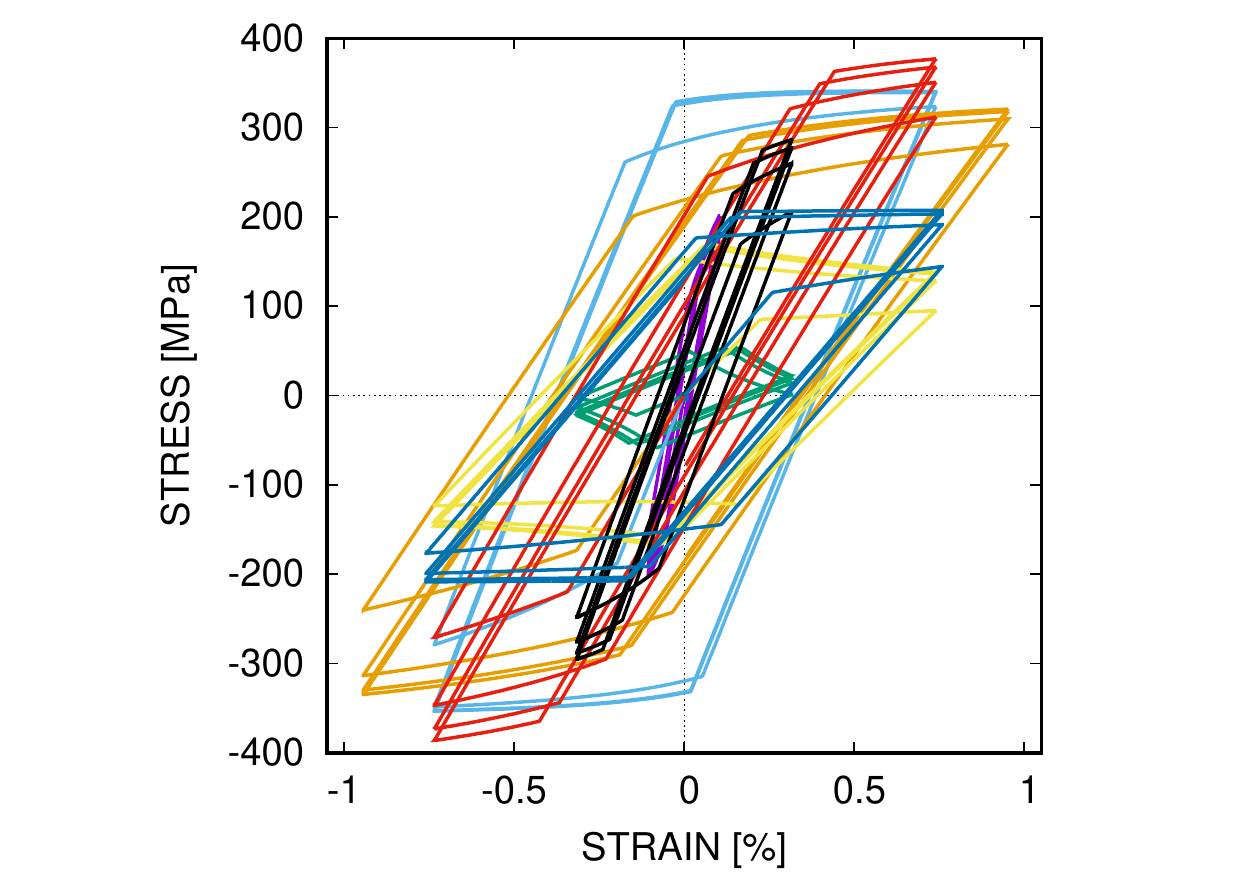}}
\caption{Stress-strain hysteresis of homogeneous UPM and J2 data for randomly selected samples across all loading modes.
Note components of the same trajectories are plotted with the same colors.
}
\label{fig:homogeneous_data}
\end{figure}

For the homogeneous material dataset used to test the \ISVNODE framework, we created response histories using three common experimental deformation modes: uniaxial, biaxial and simple shear.
The training data consists of sinusoidal displacement loading such that the stress-strain  hysteresis loops become steady after a number of cycles depending on loading frequency.
\fref{fig:homogeneous_data} shows the hysteresis data for both the UPM and J2 materials.

For each loading mode frequencies were sampled uniformly on a log scale and amplitudes were sampled uniformly on a linear scale.
For the J2 model the frequency was fixed at 0.001 Hz since it is rate independent.
The domain was a cube represented by a single element and minimum boundary condition for static determinacy were applied.
For the uniaxial mode boundary displacement of the $+x_1$ boundary was prescribed as
\begin{equation}
u_1(t) =  A_1 \sin(\omega_1 t) \eb_1 \ .
\end{equation}
For the UPM model $A_1 \in [0.1,0.5]$ and $\omega_1 \in [10^5, 10^6]$, and for the J2 model $A_1 \in [0.01,0.03]$ and $\omega_1 = 0.001$.
For the biaxial mode normal displacement on the $+x_1$ boundary was prescribed as
\begin{equation}
u_1(t) =  A_1 \sin(\omega_1 t)
\end{equation}
and on the $+x_2$ boundary
\begin{equation}
u_2(t) =  A_2 \sin(\omega_2 t) \ ,
\end{equation}
where $A_1, A_2 \in [-0.2, 0.2]$ and $\omega_1, \omega_2 \in [10^5, 10^6]$ for the UPM data, and $A_1, A_2 \in [-0.02, 0.02]$  and $\omega_2/\omega_1 \in [-1,1]$ for the J2 data.
For the simple shear mode the tangential displacement on the $+x_2$ boundary was prescribed as
\begin{equation}
u_1(t) =  A_1 \sin(\omega_1 t)
\end{equation}
where $A_1 \in [0.1, 0.2]$ and $\omega_1 \in [10^5, 10^6]$ for the UPM data, and $A_1 \in [0.005, 0.02]$  for the J2 data.
From each trajectory 4000 steps with different $\delt$ to cover the frequency range were stored and a random sample of 8000 trajectories across modes were selected for the final dataset.
These single element simulations took on average 10 cpu-ms per step (8.7 ms/step for J2 and 9.6 ms/step for UPM).
Note no shuffling of loading directions was done to promote learning, as opposed to embedding, invariance.

\subsection{Heterogeneous response} \label{sec:hetero_response}

To generate the data for samples with variable microstructure, a single set of realizations with spherical pores in a cubic sample were created and then three different material combinations were utilized to create three separate datasets.
The three material combinations were: (a) porous aluminum where the matrix followed the J2 model, (b) porous silicone where the matrix followed the UPM model, and (c) a glass-bead filled silicone were the matrix response was given by the UPM model and the inclusions were elastic.
In this last case the Young's modulus of the glass was 60 GPa and its Poisson's ratio was 0.33.
Case (c) is an example of a non-thermorheologically simple material where the similarity due to time-temperature scaling is not present.

Realizations were created by a random placement scheme of spherical voids in the sample cube  with constraints on pore overlap with other pores and the sample boundary \cite{brown2018multiscale}.
This process created unit cells with mean porosity 0.09 and standard deviation 0.03 following a beta distribution.
The 1120 realizations had porosities ranging from 0.015 to 0.017.
The cube samples were on the order of 1.5$^3$ mm$^3$  with pore radius $\approx$ 150 $\mu$m.
Pores in each of the realizations  were explicitly meshed and resulted in unstructured discretizations with 14,640 to 101,360 eight node elements.
The nominal element size was 60 $\mu$m.
Meshing was performed using the Cubit/Sculpt meshing tool \cite{owen2017hexahedral,owen2014parallel} following a process similar to Brown \etal~\cite{brown2018multiscale}.
These meshes were preprocessed into sparse adjacency matrices based on element neighbors for use with the GCNN component of the enhanced \ISVNODE, as in \cref{frankel2021mesh}.

\begin{figure}
\captionsetup[subfigure]{margin={0.5cm,0cm}}
\centering
\subfloat[J2  matrix with pores]
{\includegraphics[height=0.25\textwidth]{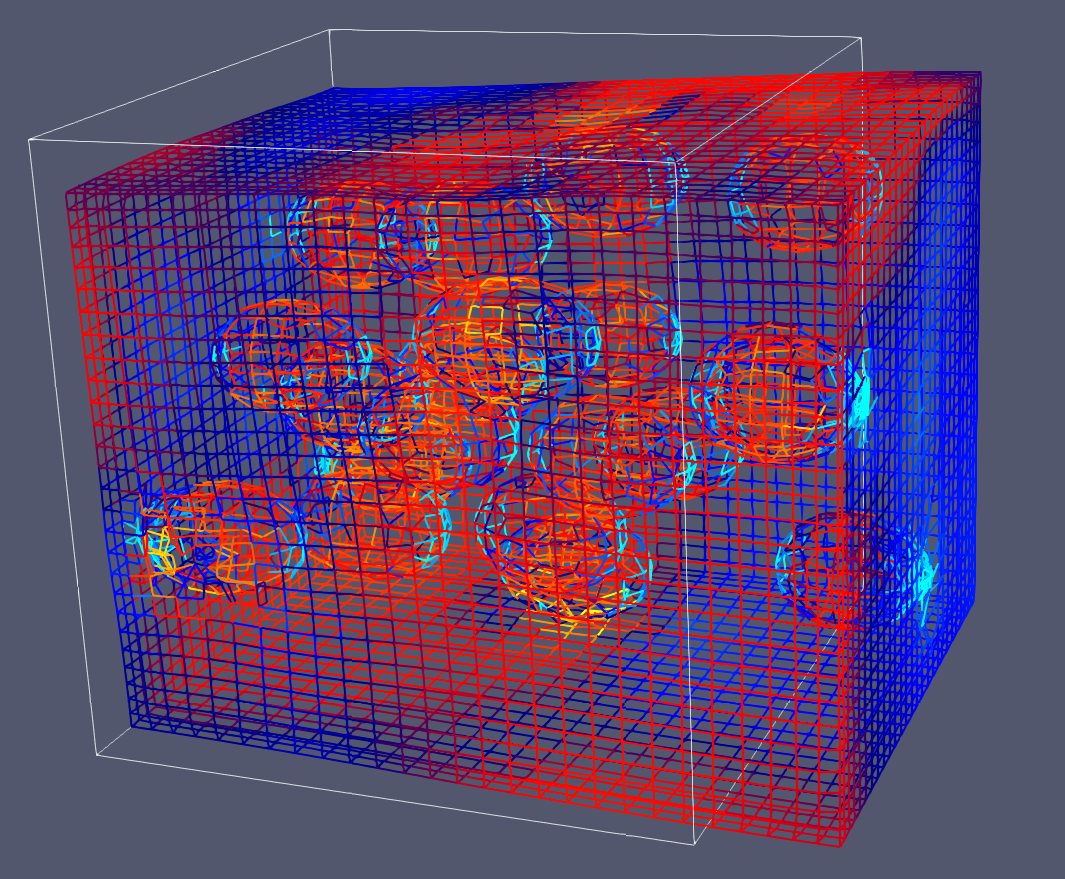}}
\subfloat[UPM matrix with pores]
{\includegraphics[height=0.25\textwidth]{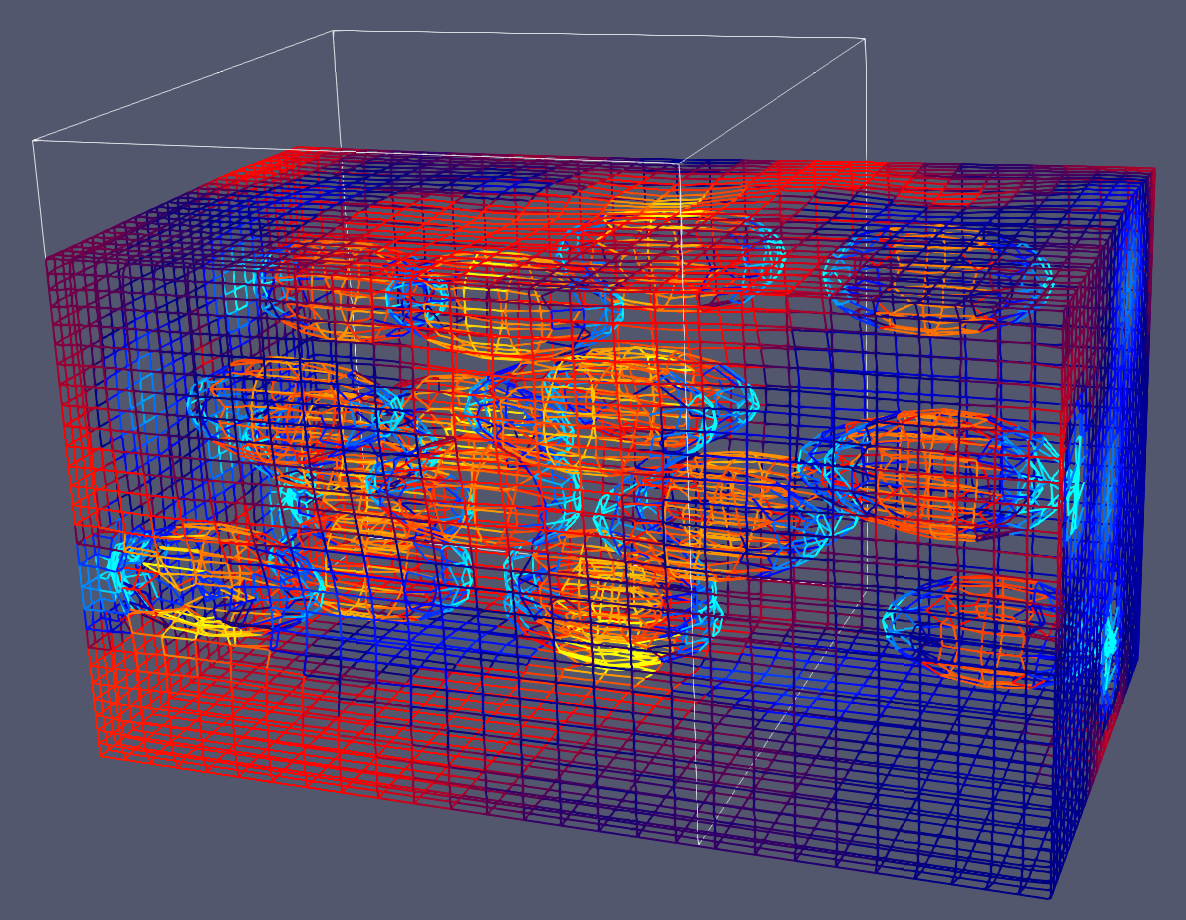}}
\subfloat[UPM matrix with elastic inclusions]
{\includegraphics[height=0.25\textwidth]{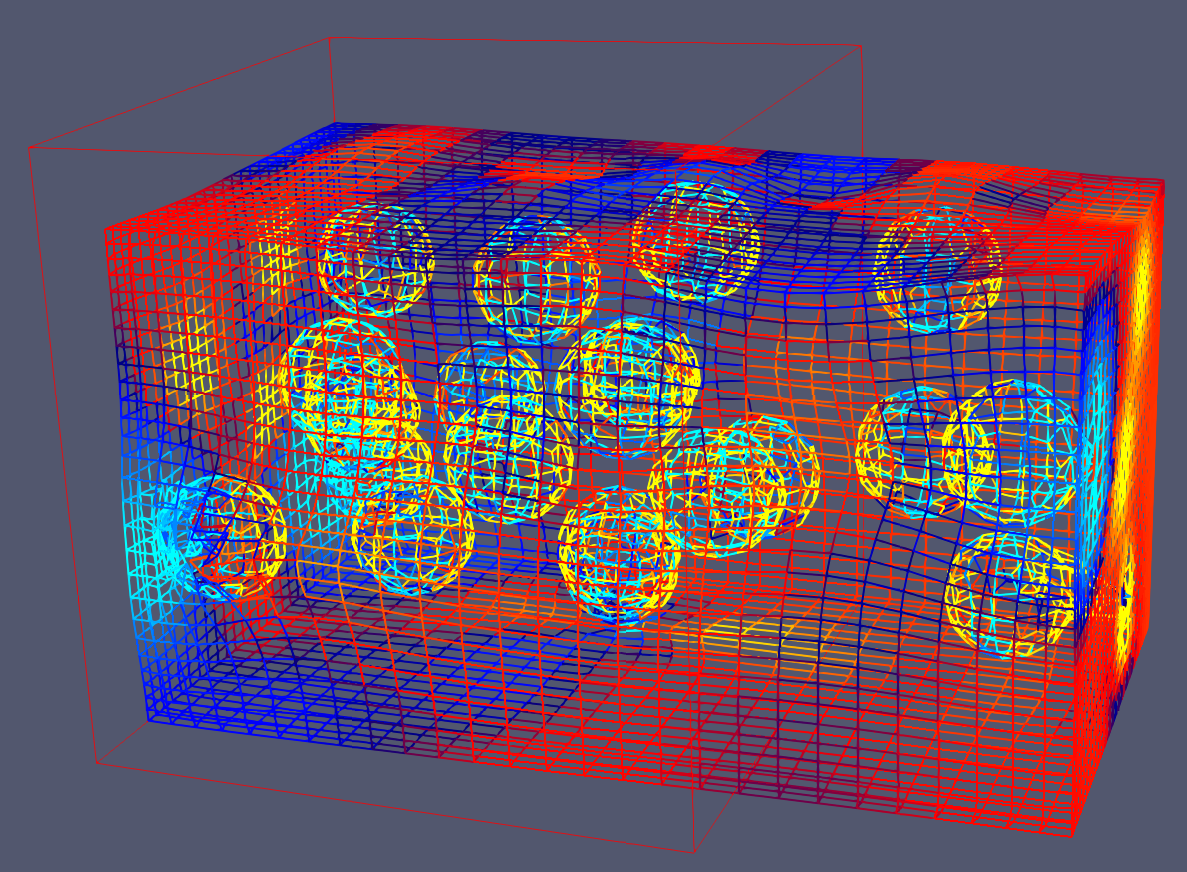}}

\caption{
Stress fields for a representative sample of the pore and inclusion data: (a) elastic-plastic J2 matrix with pores, (b) viscoelastic UPM matrix with pores, and (c) viscoelastic UPM matrix with elastic inclusions.
Mesh is colored by tensile stress, blue: $<$ 0, red: $>$ 4 MPa viscoelastic and  $>$ 700 MPa elastic-plastic.
Deformed configuration is shown at 20\% strain for the elastic-plastic simulation and at 50\% strain for the viscoelastic simulations.
The outline indicates the original configuration. }
\label{fig:pore_realizations}
\end{figure}

\begin{figure}
\centering
\subfloat[elastic-plastic J2 matrix with pores]
{\includegraphics[width=0.55\textwidth]{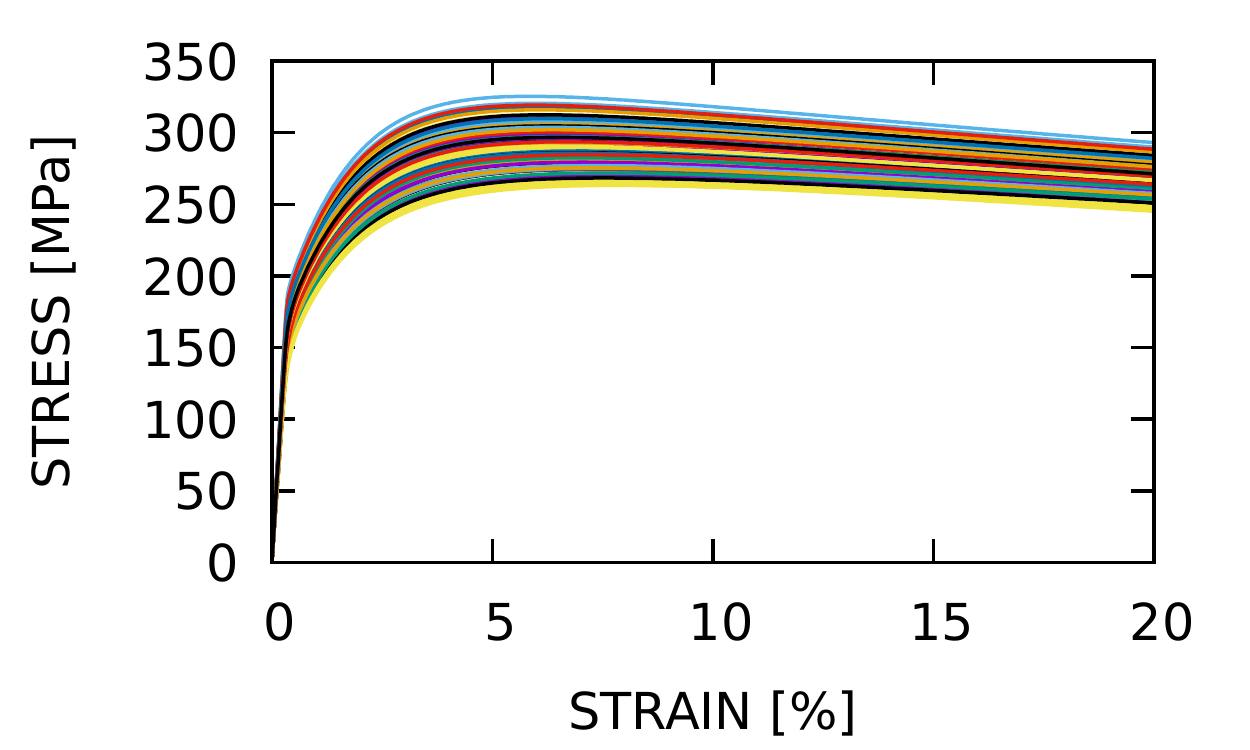}}

\subfloat[viscoelastic UPM matrix with pores]
{\includegraphics[width=0.55\textwidth]{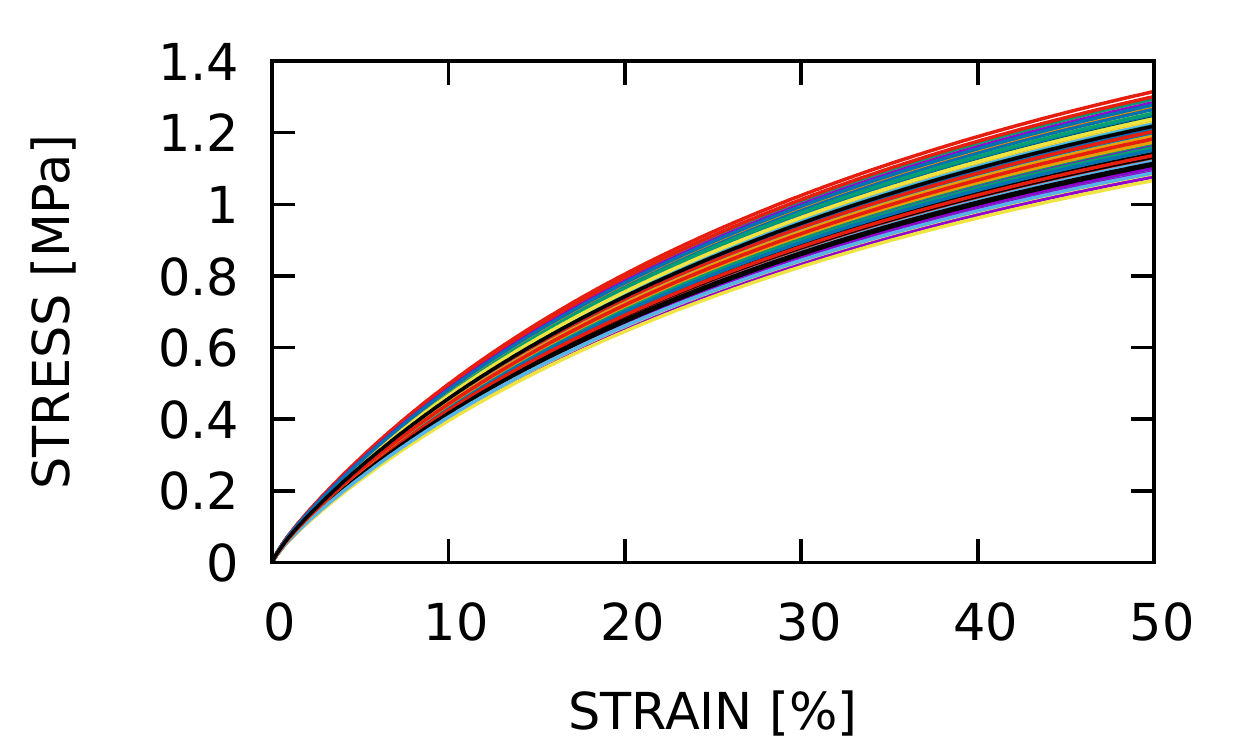}}

\subfloat[viscoelastic UPM matrix and elastic inclusions]
{\includegraphics[width=0.55\textwidth]{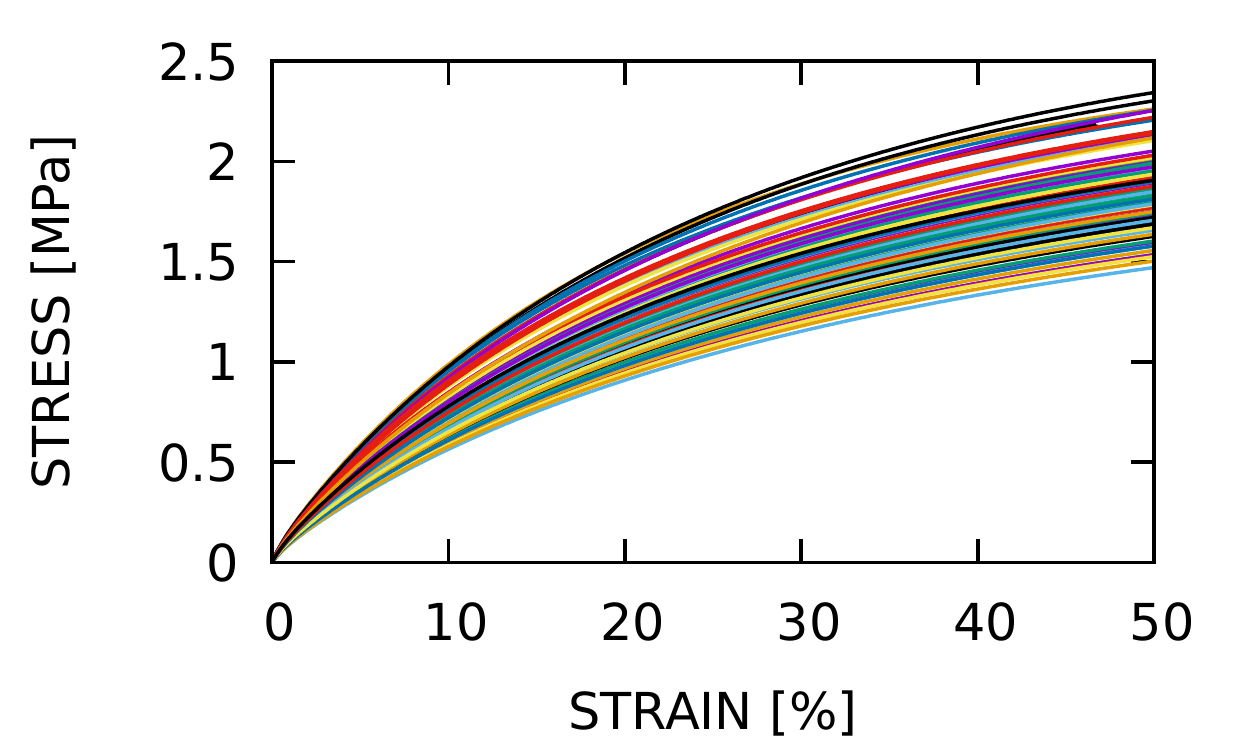}}

\caption{
Stress-strain response for the pore and inclusion data models.
Colors distinguish independent samples. }
\label{fig:pore_response}
\end{figure}

With the homogeneous models we could afford loading mode variety that was not feasible with the large meshes needed to represent the microstructure samples.
Each realization was subjected to quasi-static uniaxial tension up to 20\% engineering strain for case (a), the elastoplastic material, and to 50\% for the viscoelastic materials, cases (b) and (c).
Minimal Dirichlet boundary conditions were applied for static determinacy which resulted in inhomogeneous deformation due to the heterogeneous microstructure, as \fref{fig:pore_realizations} shows.
\fref{fig:pore_realizations} also illustrates the stress concentrations due to the second phase which are most extreme for the stiff inclusions, case (c).
Each time-step of the evolution took on the order of 10 cpu-s (10.4 s/step on average for J2, and 26.9 s/step for the UPM models).
From these simulations we extracted microstructure $\phib(\Xb)$, applied strain $\epsilon(t)$, volume-averaged tensile stress $\bar{\sigma}(t)$  data to demonstrate the efficacy of mesh-based GCNNs in the Results section.
\fref{fig:pore_response} shows the range of response for each of the three cases.
Note a simple mixture model based on the sample porosity explains some of the variance in the response but 30\% of the variance remains for case (a) after this rescaling and 45\% for case (b).
For case (c) this rescaling increases the variance.
A mixture model that takes into account the two phases would be more explanatory but would require knowing the elastic moduli of the two phases.

\section{Results} \label{sec:results}

To demonstrate the versatility of the proposed \ISVNODE network we applied it to the tasks of: (a) representing the response of homogeneous inelastic materials, and (b) the homogenization of inelastic samples with microstructure, using data described in the preceding section.

\subsection{Homogeneous material}

Using multiple loading mode data for homogeneous samples described in \sref{sec:homo_response}, we explored fundamental questions regarding of the proposed framework: (a) is one of the stress models superior to the others, (b) can an optimal latent space be ascertained, (c) how well is the dissipation inequality be satisfied, and (d) can the representation be completely trained with a reasonable amount of data and extrapolate in time.
Note that these models take approximately 15.8 $\mu$s/step to evaluate, which is approximately three orders of magnitude faster than the finite element-based data source models.

\subsubsection{Stress model formulations}
First we compare the component, tensor-basis and potential based variants
using a \ISVNODE with $N_\fb=$ 3 layers in the state-evolution model, $N_\Sb=$ 2 layers in the stress model, and $N_\ISV=$ 3 assumed states.
The component-based formulation \eref{eq:comp_stress} directly connects stress components to the outputs of certain NN nodes but lacks inherent equivariance/invariance.
The tensor-basis formulation \eref{eq:tb_stress} places part of the burden of functional complexity on the known basis used to represent the stress.
In the potential-based formulation \eref{eq:pot_stress} all information used to represent the stress is collapsed to a single, scalar output before a differentiation with respect to strain provides the stress tensor.
It is also the only formulation of the three that has direct access to the power begin dissipated.

\begin{figure}
\centering
\subfloat[true trajectories]
{\includegraphics[width=0.45\textwidth]{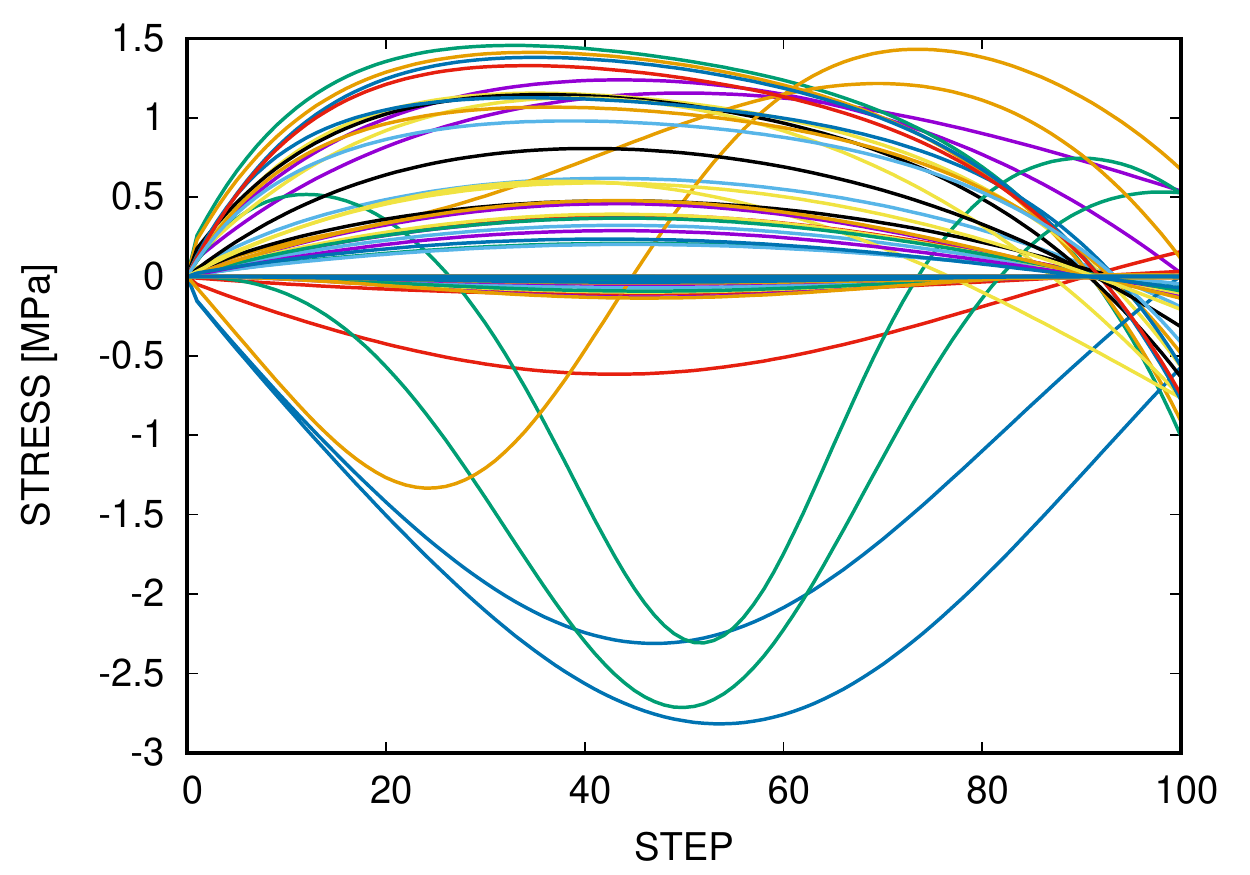}}
\subfloat[component errors]
{\includegraphics[width=0.45\textwidth]{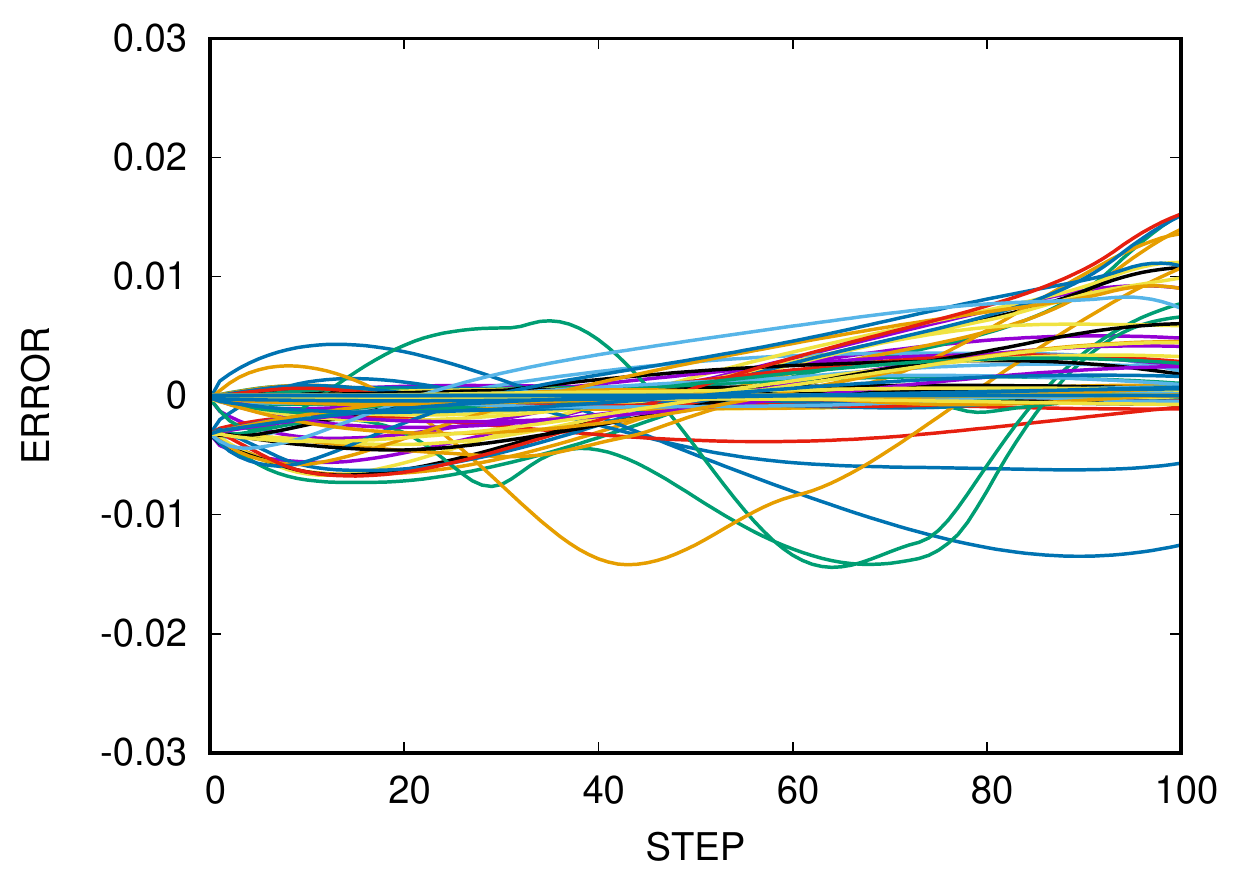}}

\subfloat[tensor basis errors]
{\includegraphics[width=0.45\textwidth]{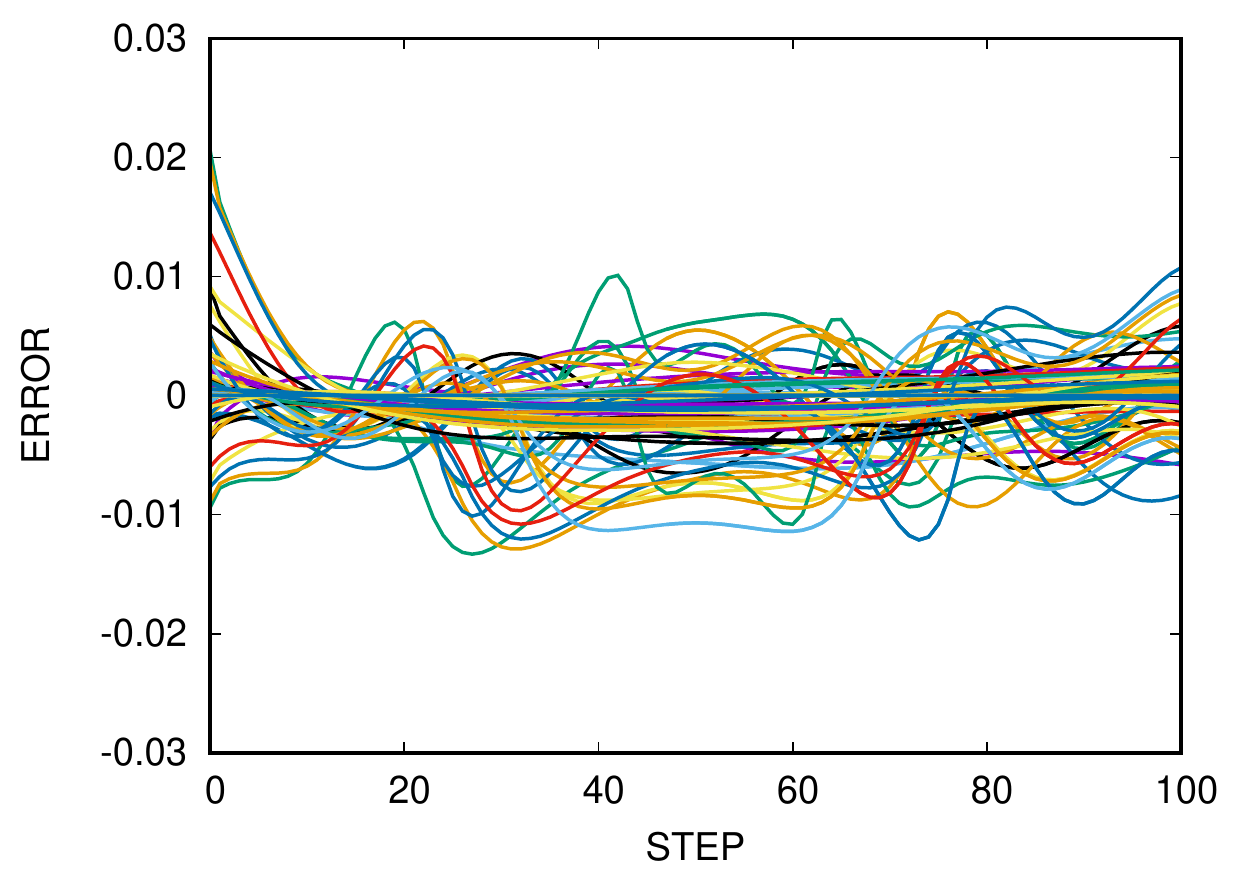}}
\subfloat[potential errors]
{\includegraphics[width=0.45\textwidth]{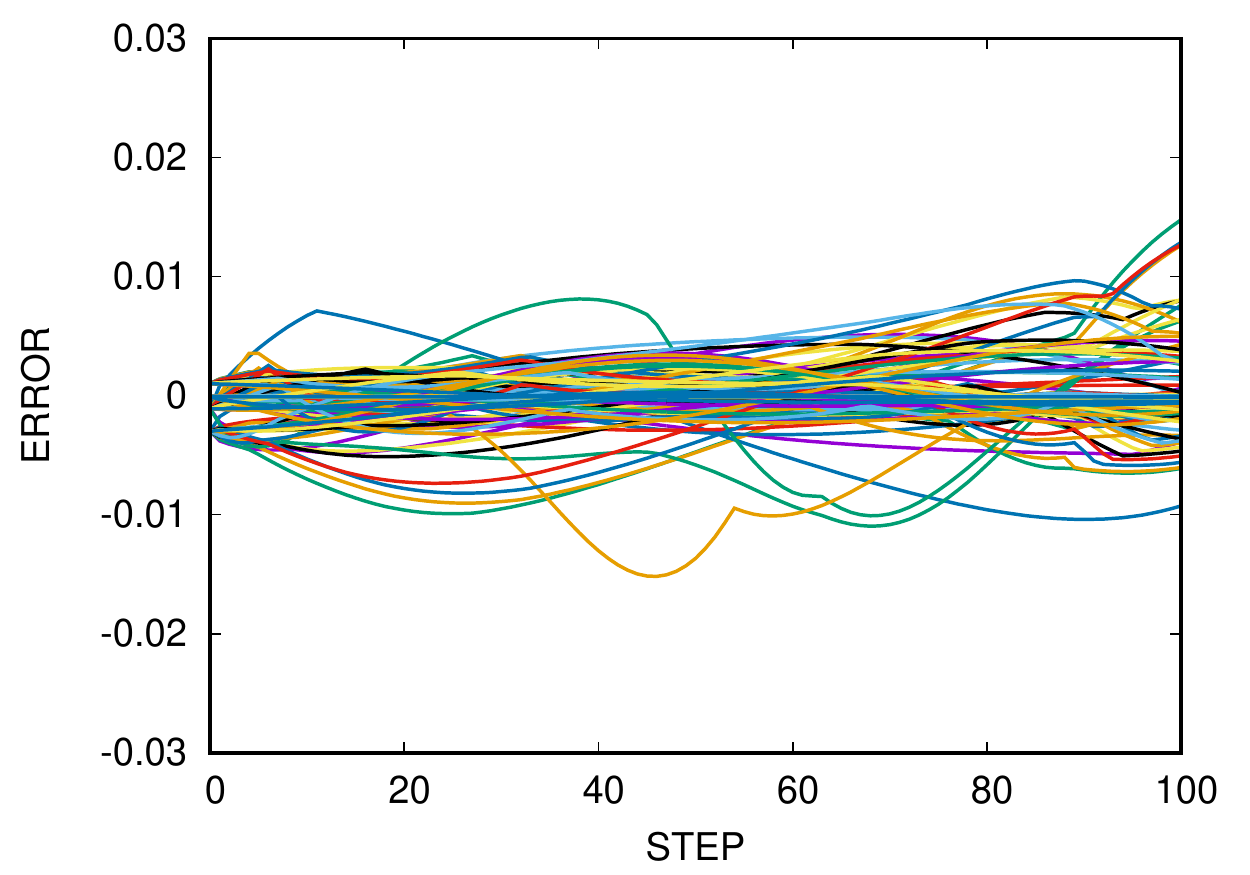}}

\caption{True stress trajectories (a) and relative trajectory errors for (b) component-based, (c) tensor basis, and (d) potential-based stress variants of the \ISVNODE.
Color distinguish different sample trajectories.
Since each sample has a distinct time-step the data shown on a step basis.
}
\label{fig:formulation_trajectories}
\end{figure}

\begin{figure}
\centering
{\includegraphics[width=0.55\textwidth]{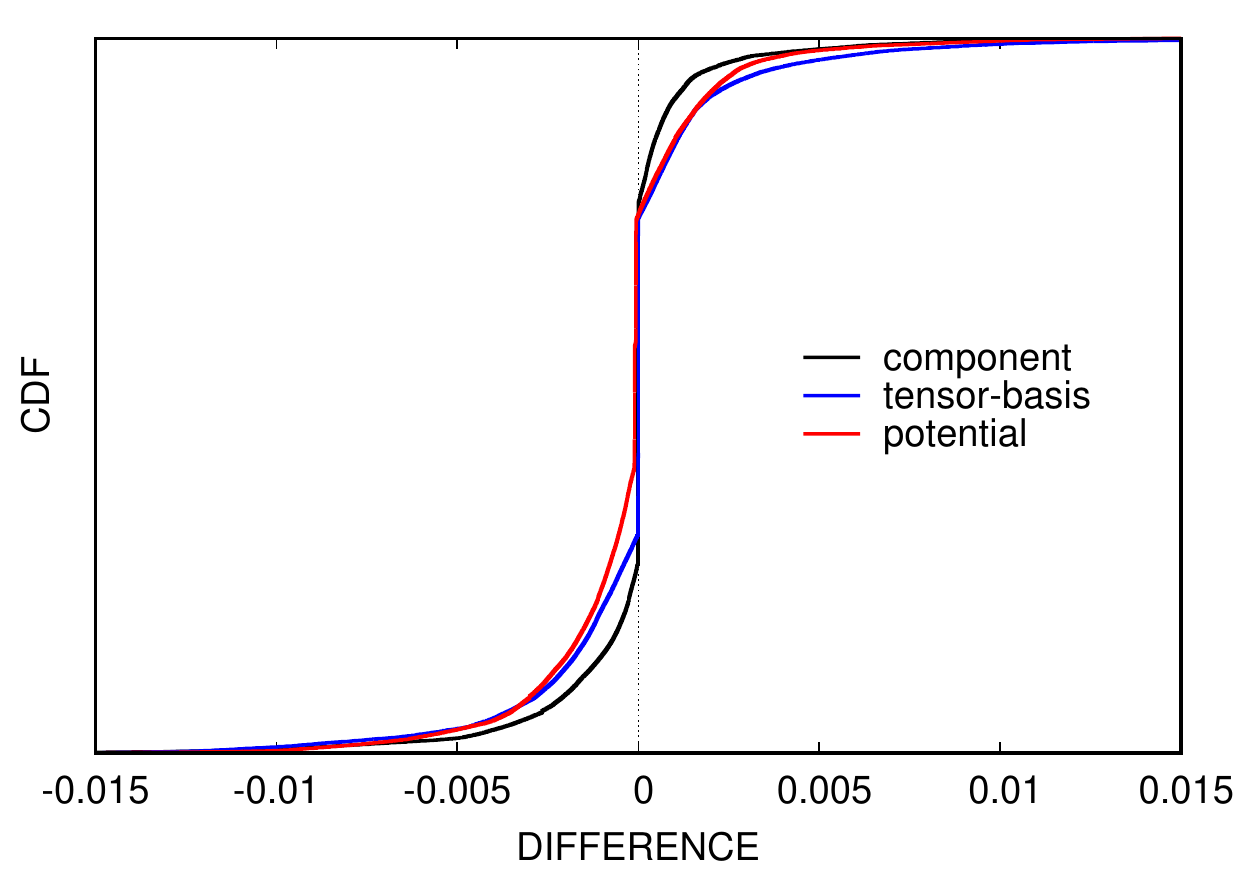}}
\caption{Comparison of CDFs of the per step errors of the three stress formulations of the \ISVNODE.
}
\label{fig:formulation_errors}
\end{figure}

\fref{fig:formulation_trajectories} shows the stress trajectories randomly selected UPM data and the relative differences of predictions for the three formulations.
Note the trajectories are plotted with respect to time-step as each trajectory has a distinct time step size and, hence, have different durations.
All formulations provide predictions of similar quality; however the errors of the potential model resemble those of the component, whereas the tensor basis errors are more oscillatory and larger at time zero.
This signature of errors is likely due to the inherent complexity of the tensor basis formulation, \ie the representation is complete to order $\Eb^2$ regardless of whether the response needs higher order basis elements.
Also the potential formulation errors have some noticeable kinks with respect to smooth data, which may be due to using relatively small MLPs with ELUs in all the \ISVNODE models.
The empirical cumulative distribution function (CDF) of the differences between the predictions and the data, shown in \fref{fig:formulation_errors}, provides a more quantitative and global view which indicates that the models are comparably accurate.
It also appears that the component based approach has slightly better accuracy over this dataset, which is understandable since loading modes were not shuffled/augmented to teach invariance and, hence, the component model could specialize to this data.
It follows that the errors for the component model would be larger if trajectories with identical invariants but distinct components were included in the validation set \cite{frankel2020tensor}.
These results suggest that each formulation can be comparably accurate; however, the potential-based has the advantage of direct access to dissipation, as will be explored in the following sections.

\subsubsection{Latent state space size}

\begin{figure}
\centering
{\includegraphics[width=0.6\textwidth]{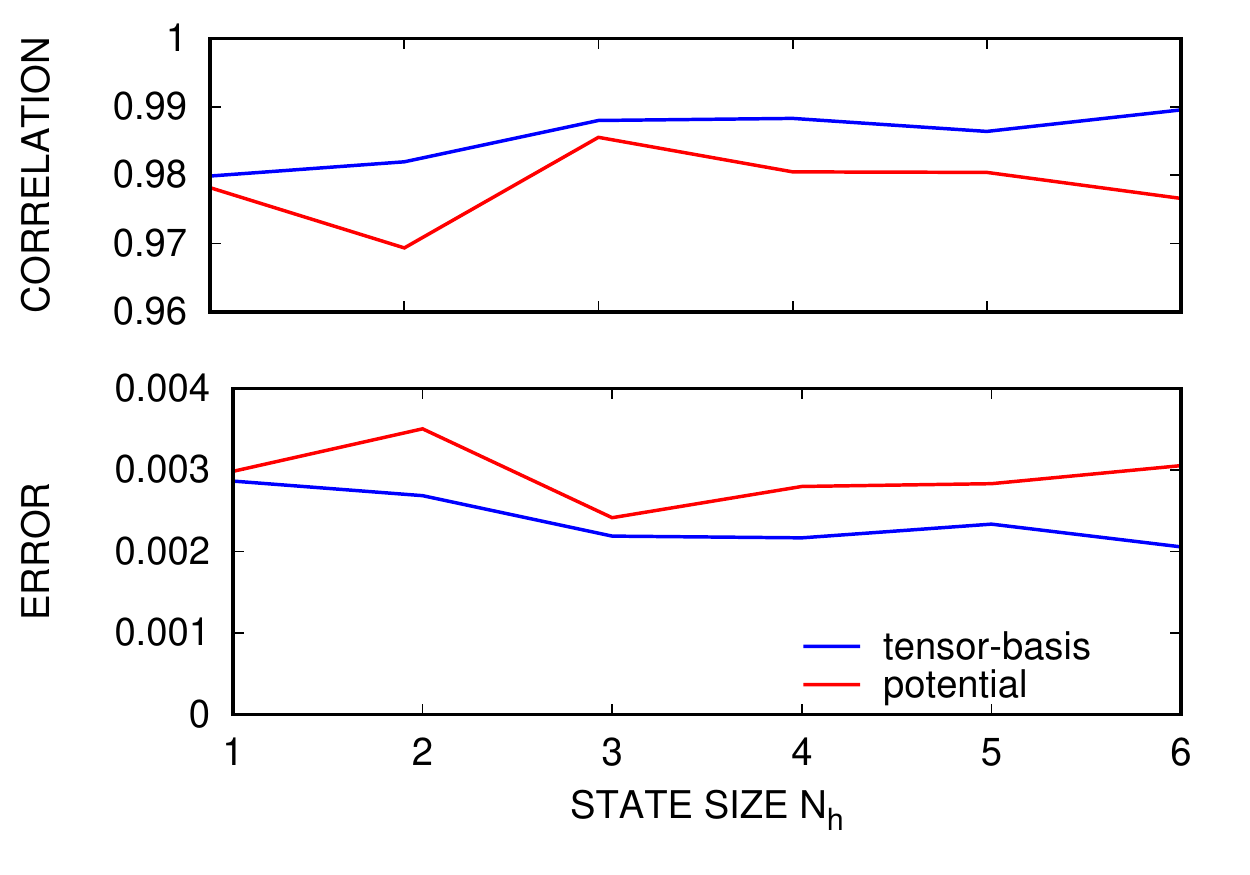}}
\caption{Sensitivity of the RMSE error and correlation to state space size $N_\ISV$.
}
\label{fig:state_size}
\end{figure}

We investigated whether there is an optimal number of internal state variables, $N_\ISV$, to describe the state space. $N_\ISV$, of a particular material response.
The existence of an optimal number of variables is plausible for actual material response and seemingly certain for data generated from a traditional parameterized model.
Using the UPM data, we varied the size of the state space $N_\ISV$ and computed the validation error of the predictions.
\fref{fig:state_size} shows the accuracy of both the tensor basis and potential-based formulations for increasing size of the inferred state space  $N_\ISV$.
It appears that the potential formulation has an optimum at $N_\ISV =$ 3, which is the number of isotropic invariants for an additional strain-like quantity, while the tensor basis formulation is generally improving with increasing state space size  for $N_\ISV \le 6$.
The weak sensitivity to the size of the state space is likely due the inherent quality of NN to handle redundancy by creating correlated outputs, as was observed in \cref{frankel2019predicting} where correlated features were extracted from images when more output nodes than were needed were available to the NN.
This topic will be revisited with the J2 data in \sref{sec:j2_nn}.

\subsubsection{Hidden variable evolution and dissipation}
We devised two constraints to ensure proper dissipation.
For the potential-based variant, we have direct access to the power expended by the hidden states and a connection to the second law.
This connection is encapsulated by the inequality constraint \eref{eq:pot_penalty}.
Without this direct connection, we resorted to matching the power expended by the model to that of the data using the equality constraint \eref{eq:tb_penalty}.

\begin{figure}
\centering
{\includegraphics[width=0.6\textwidth]{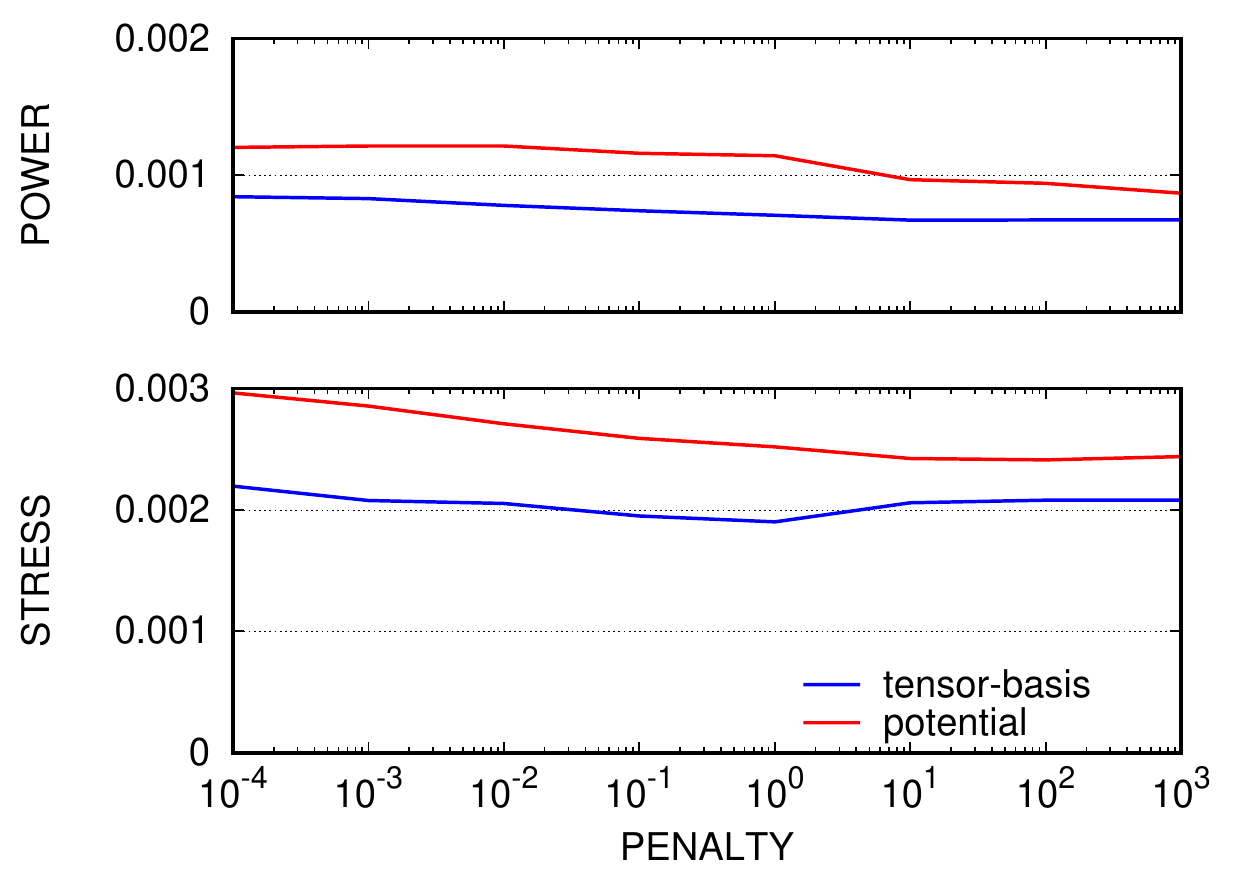}}
\caption{Sensitivity of the stress and power RMSE errors to the penalty parameter $\varepsilon$.
}
\label{fig:dissipation_penalty}
\end{figure}

\fref{fig:dissipation_penalty} shows the error in the stress \eref{eq:loss} and the error in the total power \eref{eq:tb_penalty} for the two formulations with $N_\ISV = 3$ and the UPM data.
Clearly there is an optimal value for the penalty for both formulations based on the stress error, but their accuracies are both relatively insensitive to the particular value of the penalty $\varepsilon$.
This indicates that the expended power error $\| (\Sb - \hat{\Sb}) \cdot \dot{\Eb}\|$, which is essentially the stress error in the direction of the strain rate $\dot{\Eb}$, is generally compatible with the formulations.
In both cases the power error keeps decreasing with increased penalty $\varepsilon$, which is expected.
Note for the tensor basis formulation, this error is exactly what is being penalized, whereas for the potential formulation  any positive internal state power   (signaling second law violations and instability) is penalized.

\begin{figure}
\centering
\subfloat[state $\epsilon = 0$]
{\includegraphics[width=0.45\textwidth]{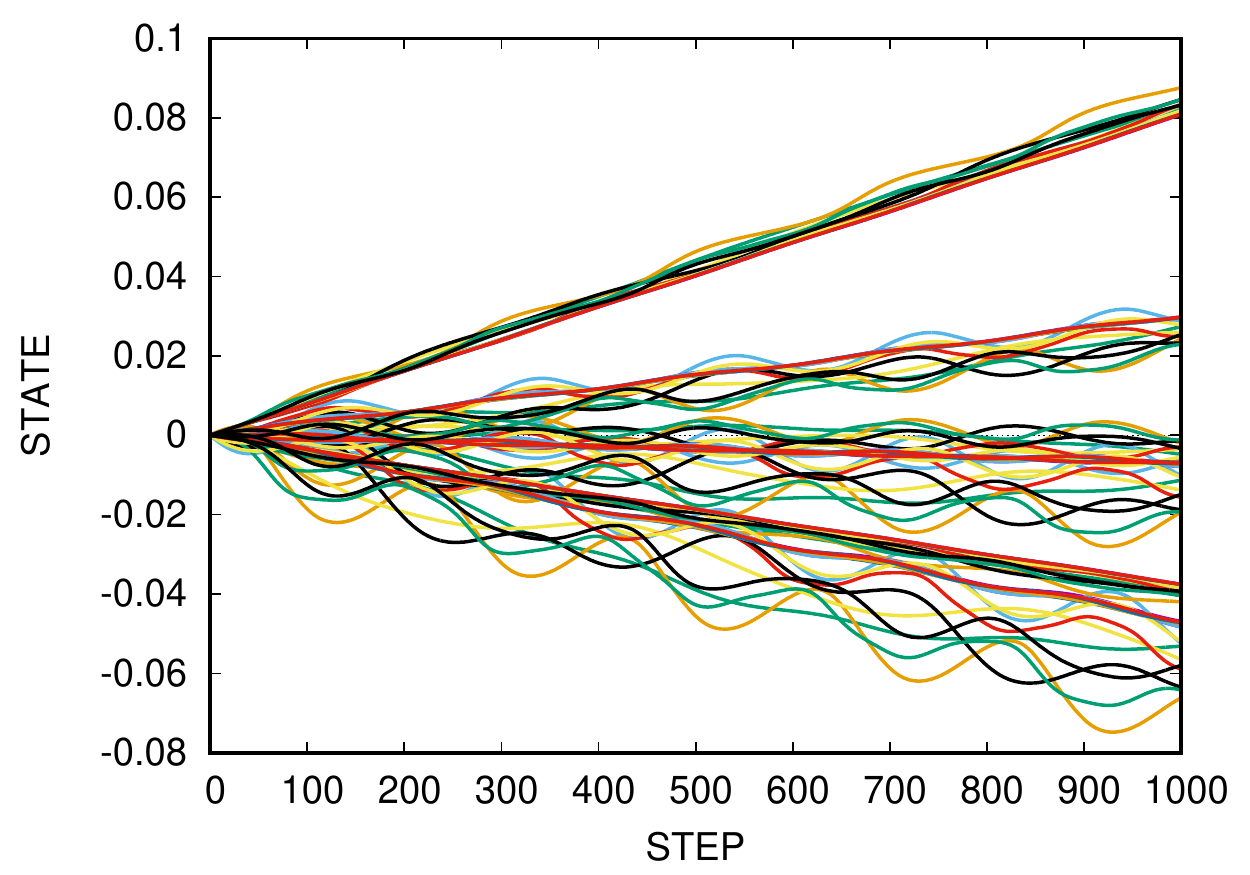}}
\subfloat[state $\epsilon = 10^3$]
{\includegraphics[width=0.45\textwidth]{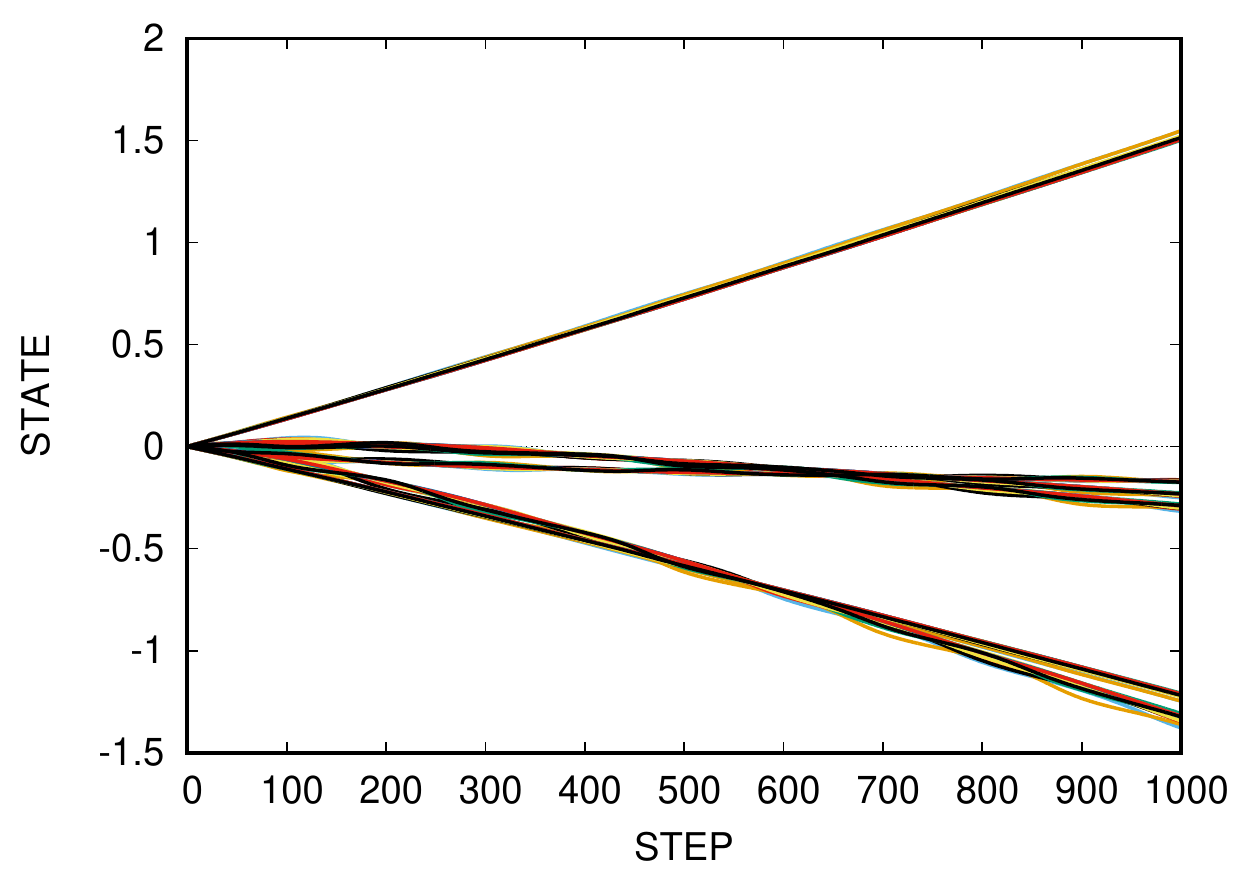}}

\subfloat[power $\epsilon = 0$]
{\includegraphics[width=0.45\textwidth]{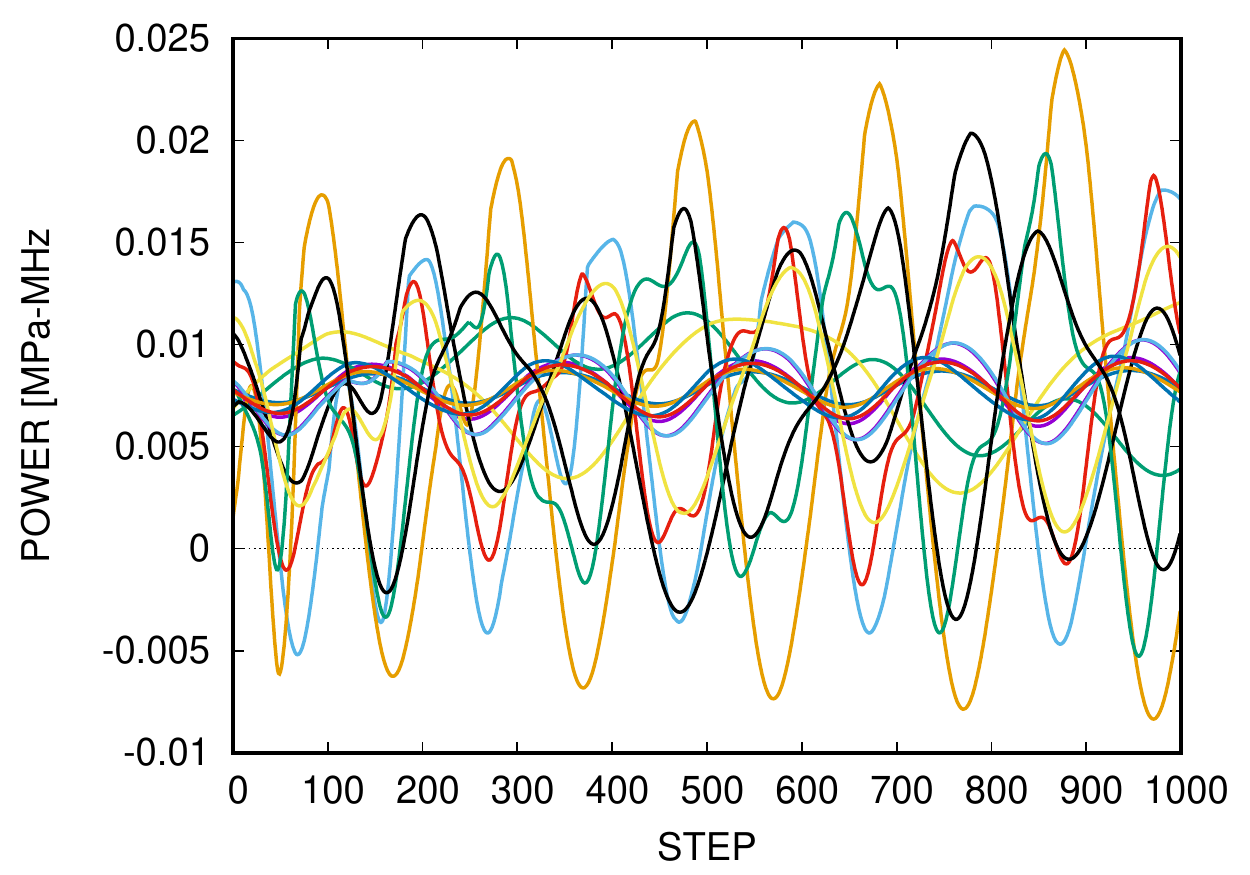}}
\subfloat[power $\epsilon = 10^2$]
{\includegraphics[width=0.45\textwidth]{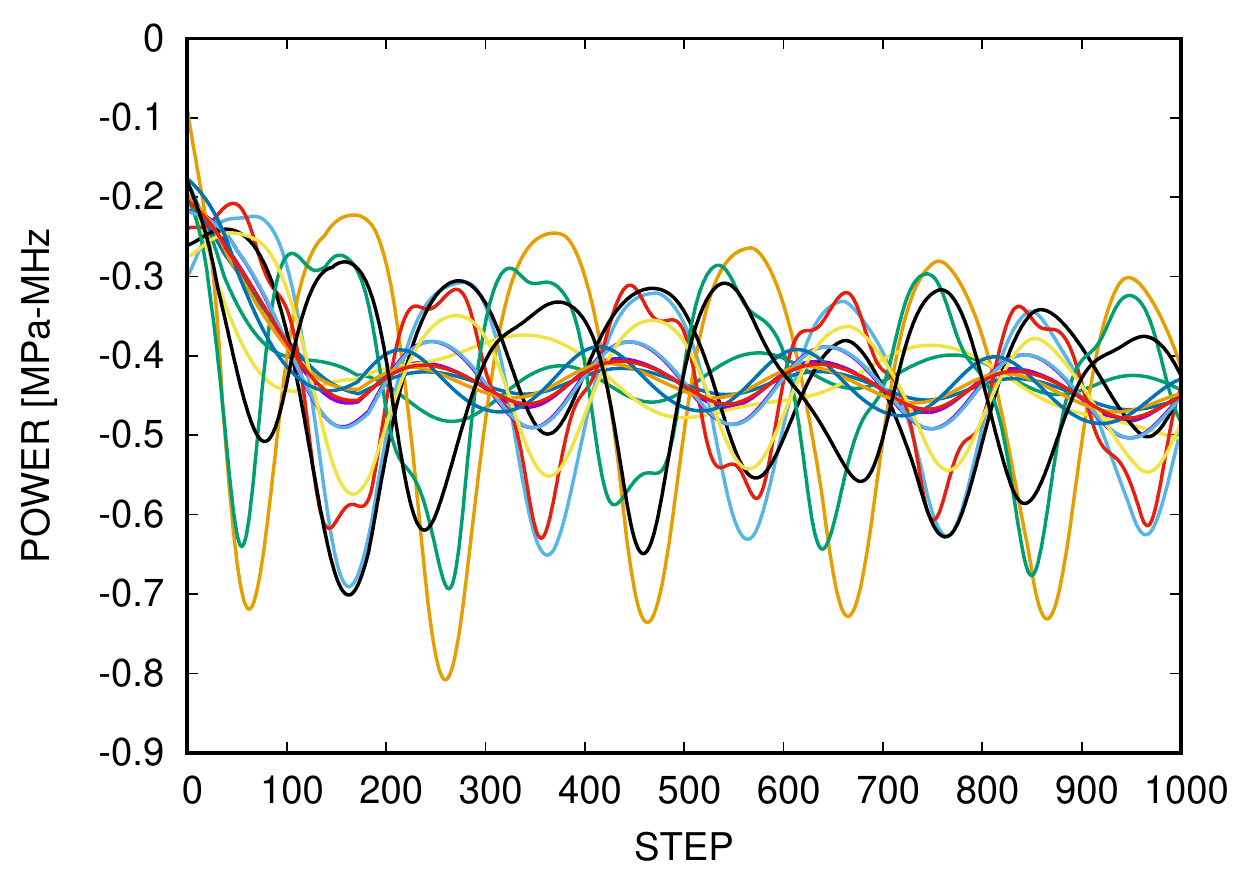}}
\caption{State evolution and power without ($\varepsilon = 0$) and with ($\varepsilon > 0$) dissipation constraint for potential formulation.
State trajectories from same sample are plotted with the same color.
}
\label{fig:state_evolution}
\end{figure}

For potential formulation we examined the effect of the penalization on the internal evolution.
\fref{fig:state_evolution} shows the state evolution and the internal state power with and without penalization.
Recall that the timestep for each trajectory is distinct.
Clearly the penalization is necessary to prevent positive internal state power and second law violations.
Other than guiding the optimal weights during training, the penalization of non-dissipation has no other effect on predictions shown.
In fact penalization results in the power expended to increase over a few cycles of displacement loading and then level off as the material reaches a steady state with the loading, which is expected from a viscoelastic material with harmonic loading.
For both cases the states follow generally linear growth trends with sinusoidal fluctuations.
There is correlation with the state fluctuations and the power expended, as expected.
Penalization also has the effect of reducing the oscillations relative to the linear trends in the state evolution for the harmonically forced dataset.
For this material there is always dissipation with loading so we do not expect to observe phases where $\dot{\ISV} = 0$.
This will be revisited in \sref{sec:j2_nn}.

\subsubsection{Time extrapolation}

\begin{figure}
\centering
\subfloat[error]
{\includegraphics[width=0.45\textwidth]{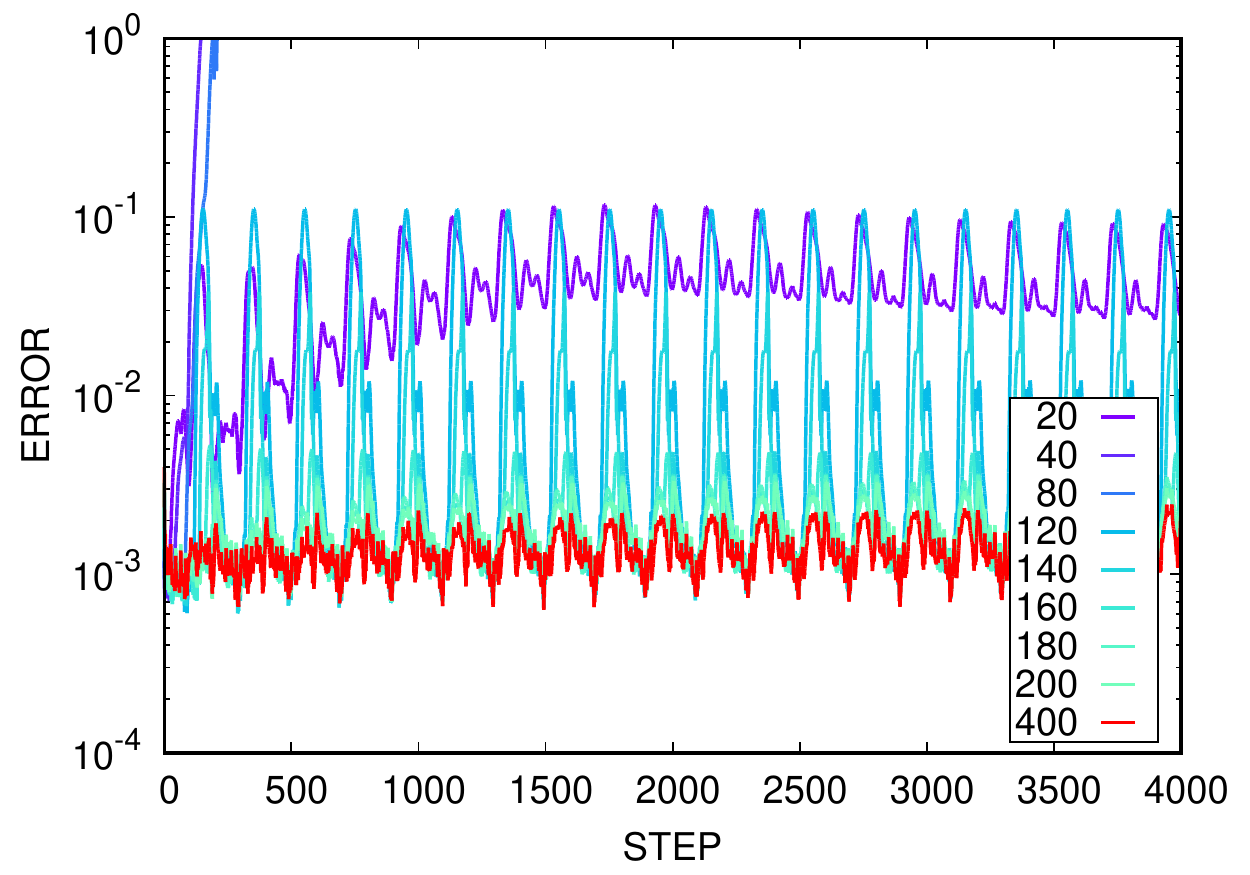}}
\subfloat[correlation]
{\includegraphics[width=0.45\textwidth]{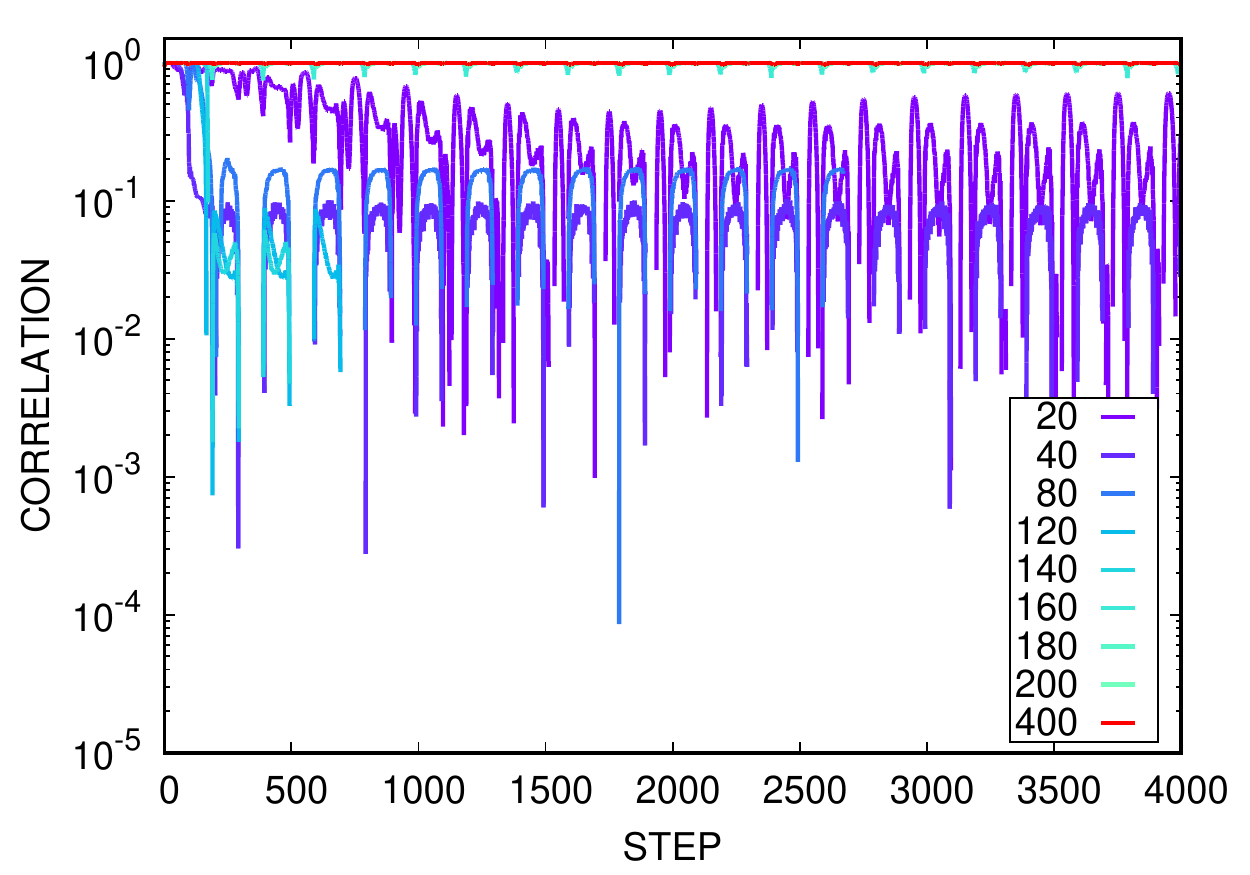}}
\caption{Convergence of the RMSE error and correlation with sequential training.
The RMSE and correlation of predictions to time step 4000 are plotted for a sequence of trainings where only $N^i_\text{steps}=20,40,\ldots,400$ of the data is used in the training.
The predictions are extrapolations in time since $N^i_\text{steps} \ll 4000$.
}
\label{fig:time_extrapolation}
\end{figure}

As a last investigation with the homogeneous UPM material dataset, we tested whether training eventually becomes complete, in the sense the extrapolation errors decrease, with increasing duration of the training data.
\fref{fig:time_extrapolation} shows convergence of time extrapolation errors with increasing training steps $N_\text{steps}$.
Clearly the extrapolation errors become effectively constant in time indicating that the essentials of the material dissipation and internal state evolution have been incorporated into the \ISVNODE model.
Ultimately the accuracy of the \ISVNODE is limited by the complexity of the flow and stress MLPs.
Results are shown for the potential based \ISVNODE variant; those for the tensor basis formulation are similar.

\subsubsection{Conservation and dissipation} \label{sec:j2_nn}
The \ISVNODE model, in particular the potential-based formulation, has the ability to represent both dissipative processes, where the hidden state evolves, and conservative processes, where the hidden state should be constant and the stress potential is only a function of the varying strain.
The J2 elastic-plastic data has both aspects, hysteresis loops which begin elastic and then transition to plastic flow when yield is encountered.

\begin{figure}
\centering
{\includegraphics[width=0.50\textwidth]{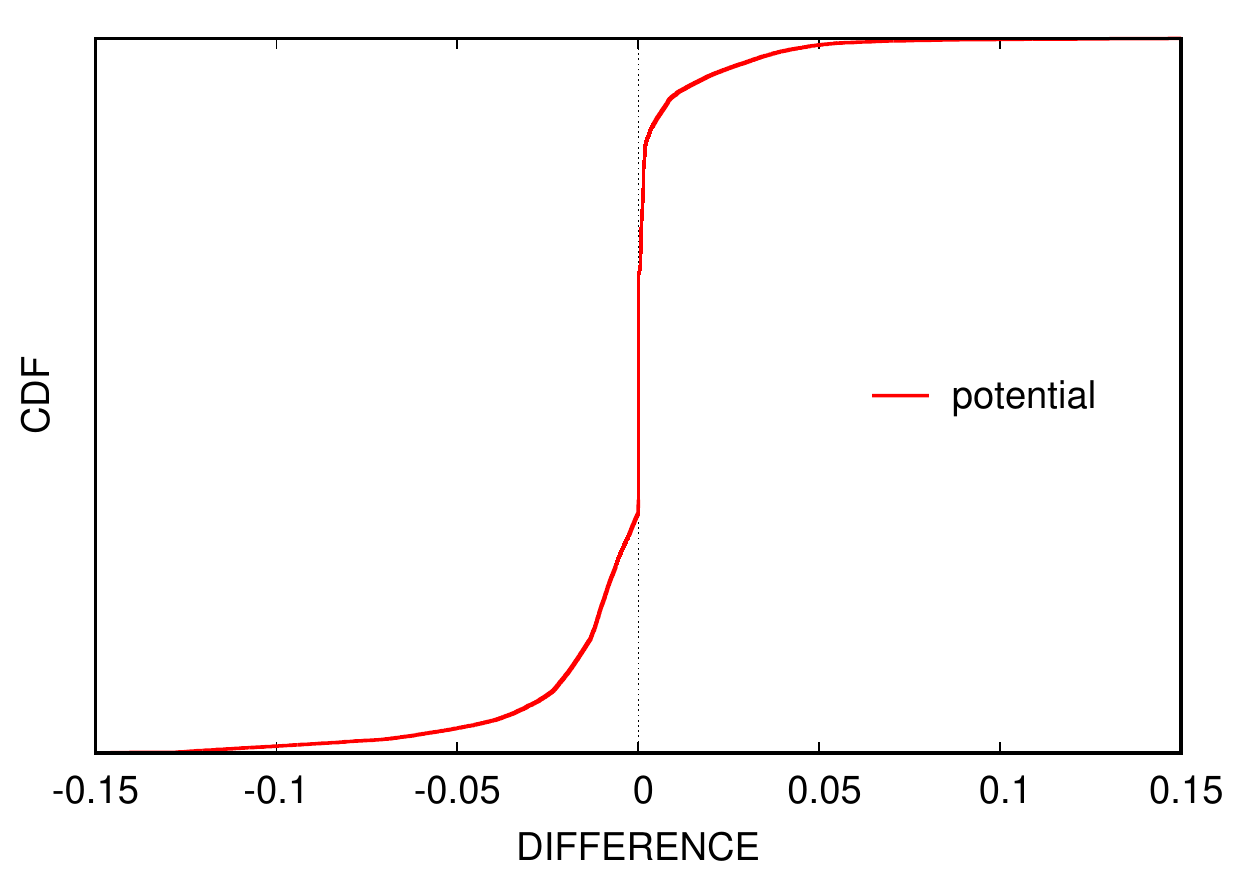}}
\caption{CDF of the differences between the true and predicted trajectories of potential-based \ISVNODE trained to J2 data.
}
\label{fig:j2_cdf}
\end{figure}

Using a \ISVNODE with $N_\fb =3$ layers in the flow model and $N_\Sb =3$ layers in the stress model, we first determined an optimal number of state variables $N_\ISV$.
The \ISVNODE model of the J2 had similar trends but stronger dependence on the number of state variables than the \ISVNODE model of the UPM data.
It appears that the more complex (higher $N_\ISV$) models train slower.
For a feasible number of sequential training stages and epochs, models with $N_\ISV =$ 1,2 are comparably accurate over the initial part of the trajectories.
\fref{fig:j2_cdf} shows the CDF of prediction errors, which are distinctly higher than for UPM data.
This is expected due to the difficulty of representing the non-smooth J2 data with two smooth functions $\fb$ and $\Sb$.

\begin{figure}
\centering
{\includegraphics[width=0.95\textwidth]{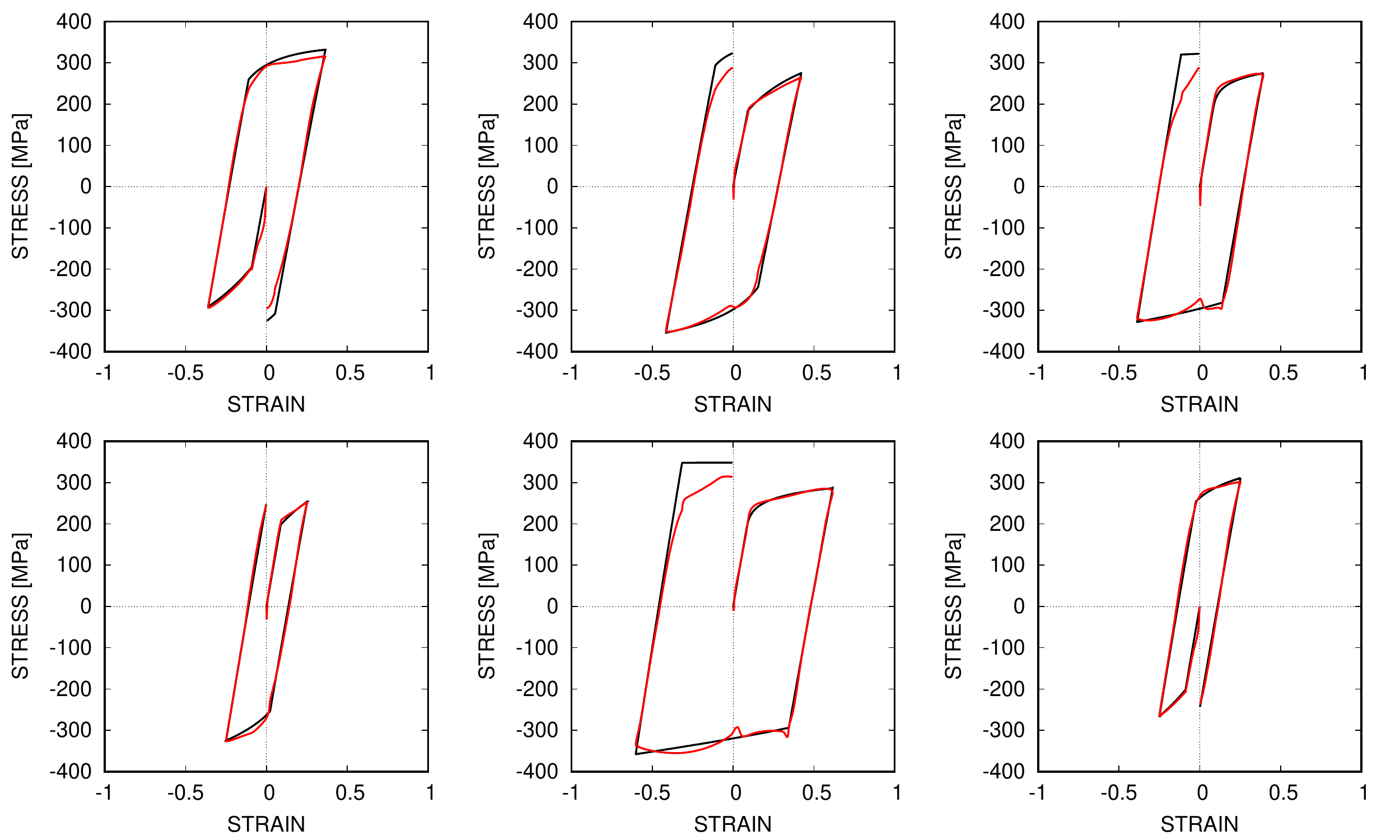}}
\caption{Comparison of true (black) and predicted (red) stress-strain hysteresis for the J2 data.
}
\label{fig:j2_comp}
\end{figure}

\begin{figure}
\centering
{\includegraphics[width=0.6\textwidth]{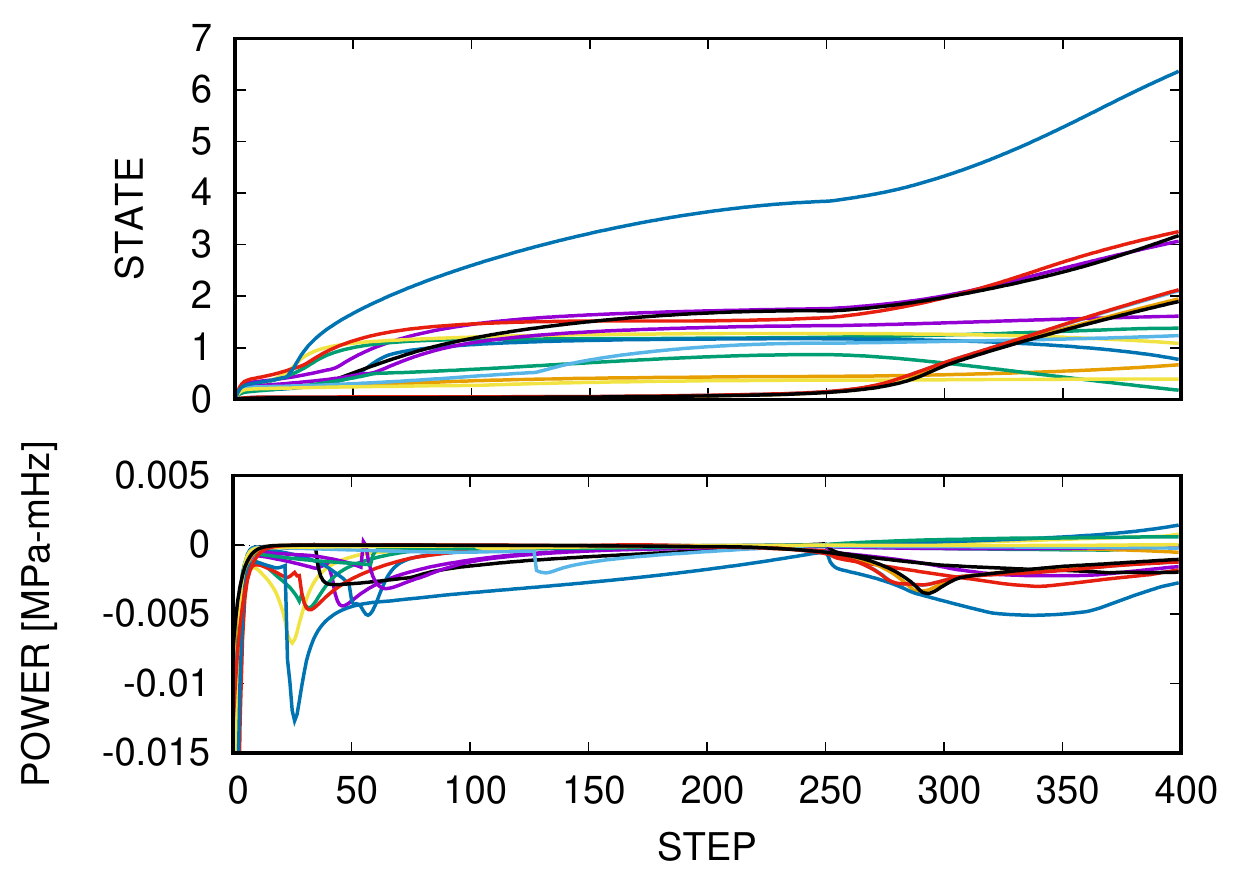}}
\caption{State evolution for potential-based \ISVNODE of the J2 data.
}
\label{fig:j2_state}
\end{figure}

\fref{fig:j2_comp} shows the true and predicted stress-strain hysteresis for an \ISVNODE with $N_\ISV = 2$ and trained with penalty $\varepsilon = 10.0$.
The \ISVNODE predictions are close to the data even through multiple reversals.
The corresponding state evolution and state power are shown in \fref{fig:j2_state}.
Consistent with the expectations from ISV theory, there are stages were the states are essentially constant, corresponding to elastic formation, and others were there is dissipation.
Given the smoothness of the underlying MLPs in the \ISVNODE, the switching between dissipation and conservation is a smooth approximation of the abrupt transition in the data.
Also noticeable in \fref{fig:j2_comp} are spurious oscillations when crossing zero strain with a negative strain rate on a few of the trajectories; we believe this is due to insufficient sampling of this regime, although it also could be the limited number of nodes in the MLPs since similar kinks are apparent in \fref{fig:formulation_trajectories}.
Nevertheless the accuracy of the trajectories appears to recover before degrading after multiple reversals.
Clearly there are limitations with this representation of the J2 data, that being said experimental data tends to be generally smoother than the response of the  J2 model.
This topic will be discussed further in \sref{sec:conclusion}.

\subsection{Homogenized microstructure}

To model the volume-averaged stress response of heterogeneous samples we used a \ISVNODE with $N_\fb=$ 3 layers in the state-evolution model, $N_\Sb=$ 2 layers in the stress model including the linear output layer and $N_\ISV=$ 8 assumed states.
A larger number of internal states was selected based on the assumption that the internal state of a heterogeneous sample will be more complex; however, no tuning of $N_\ISV$ was done due to the expense of training the GCNN-\ISVNODE.
The size assumed state $N_\ISV$ was augmented by the output $\varphib$ of a graph convolutional NN reducing the microstructure $\phib$ to relevant features.
The graph convolutional NN subcomponent, refer to \fref{fig:heterogeneous_architecture}b, processed the binary porosity field $\varphib$ on the unstructured meshes with $N_\text{conv}=$ 4 convolutional layers each with $N_\text{filter}=$ 32.
This output was reduced by global average pooling, and finally processed by $N_\text{dense}=$ 3 densely connected layers including the linear output layer resulting in the structural features $\varphib$ correlated with the output $\Sb(t)$.
In this study we only had tensile data, so we reduced the input strains, strain rates and stress to one component.
Unlike for the main \ISVNODE, a ReLU activation was used in the GCNN.
Also, unlike our previous CNN-RNN architectures \cite{frankel2019predicting,frankel2020prediction,frankel2021mesh}, the output of reducing the initial microstructure with a GCNN is a true initial condition that augments the state space, as opposed to a constant loading-like input.
This feature makes this architecture consistent with homogenization as an initial value problem where microstructure can evolve.

\begin{figure}
\centering
{\includegraphics[width=0.6\textwidth]{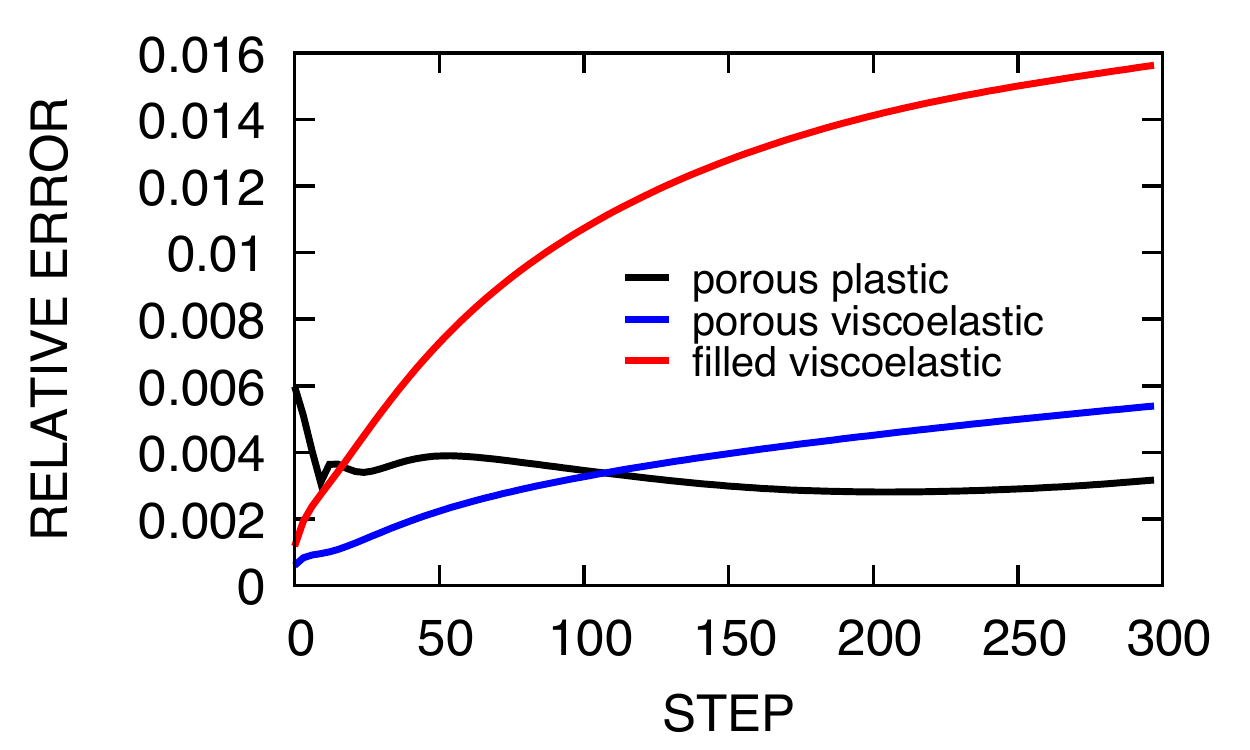}}
\caption{Evolution of mean error for \ISVNODE models of  J2 matrix with pores, UPM matrix with pores, and UPM matrix with elastic inclusions.
}
\label{fig:gcnn_comp}
\end{figure}

\begin{figure}
\centering
\subfloat[elastic-plastic matrix with pores]
{\includegraphics[width=0.45\textwidth]{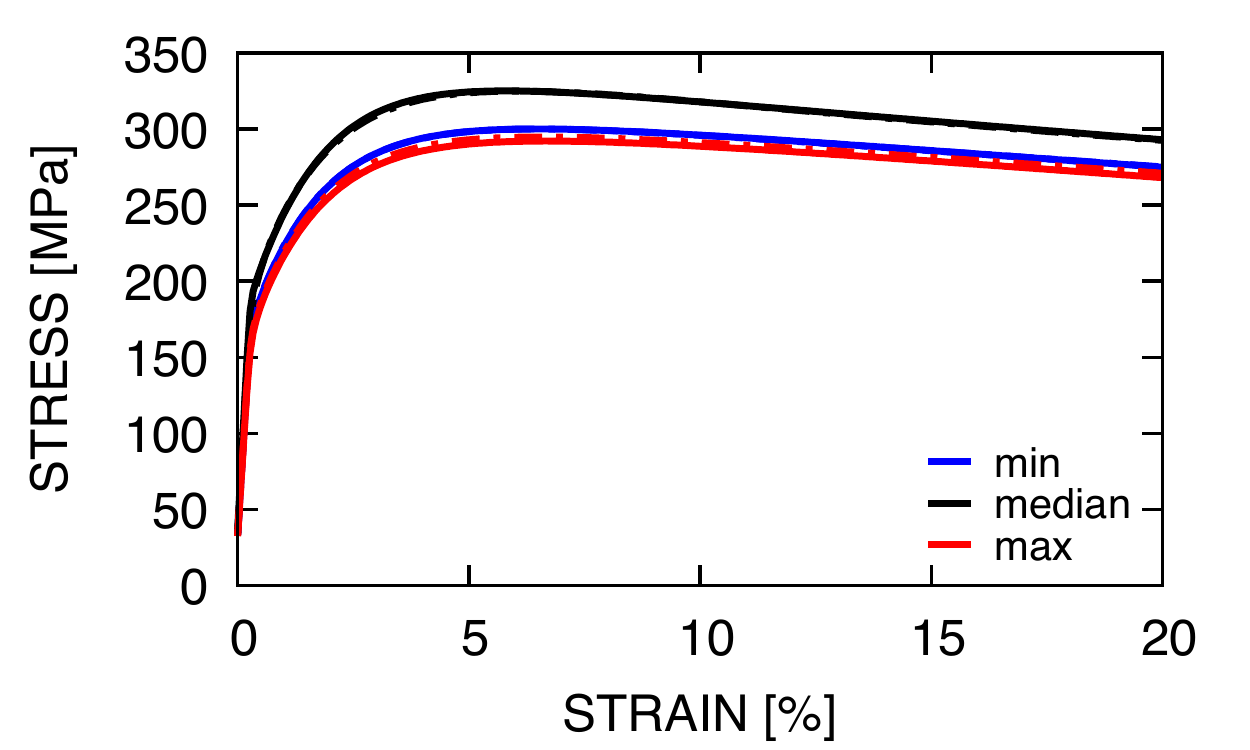}}

\subfloat[viscoelastic matrix with pores]
{\includegraphics[width=0.45\textwidth]{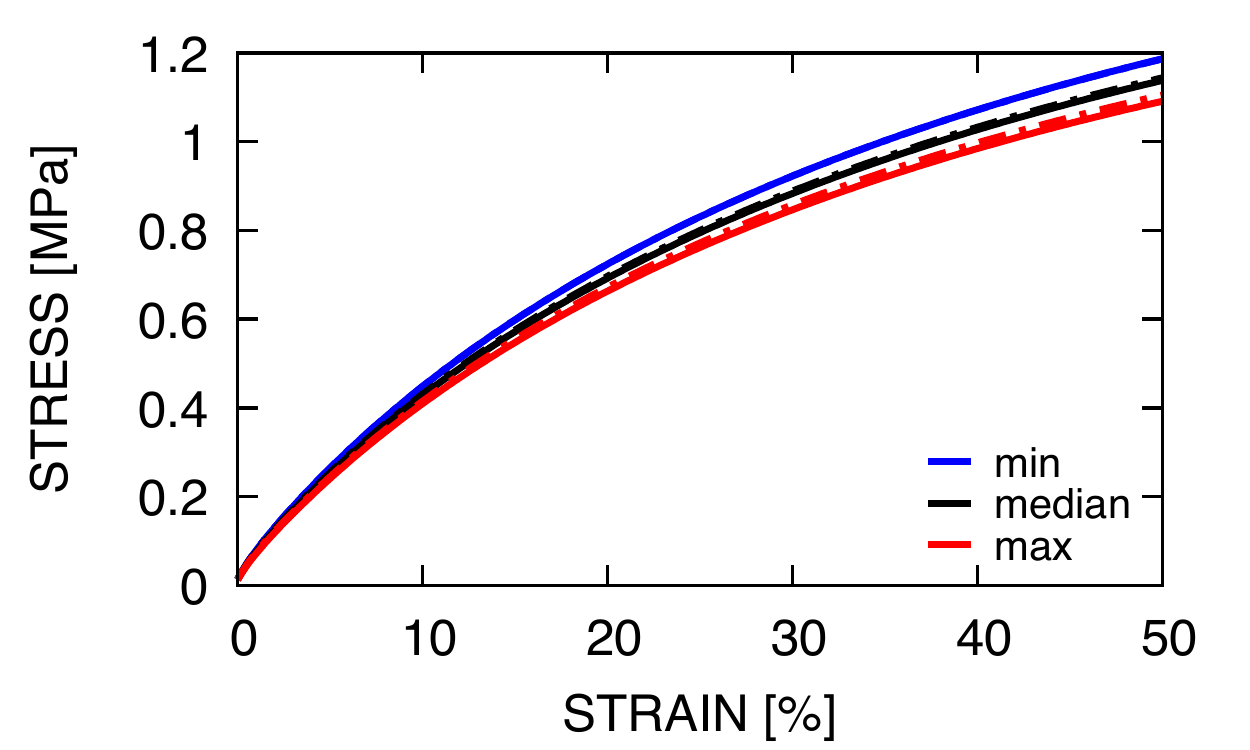}}

\subfloat[viscoelastic matrix with elastic inclusions]
{\includegraphics[width=0.45\textwidth]{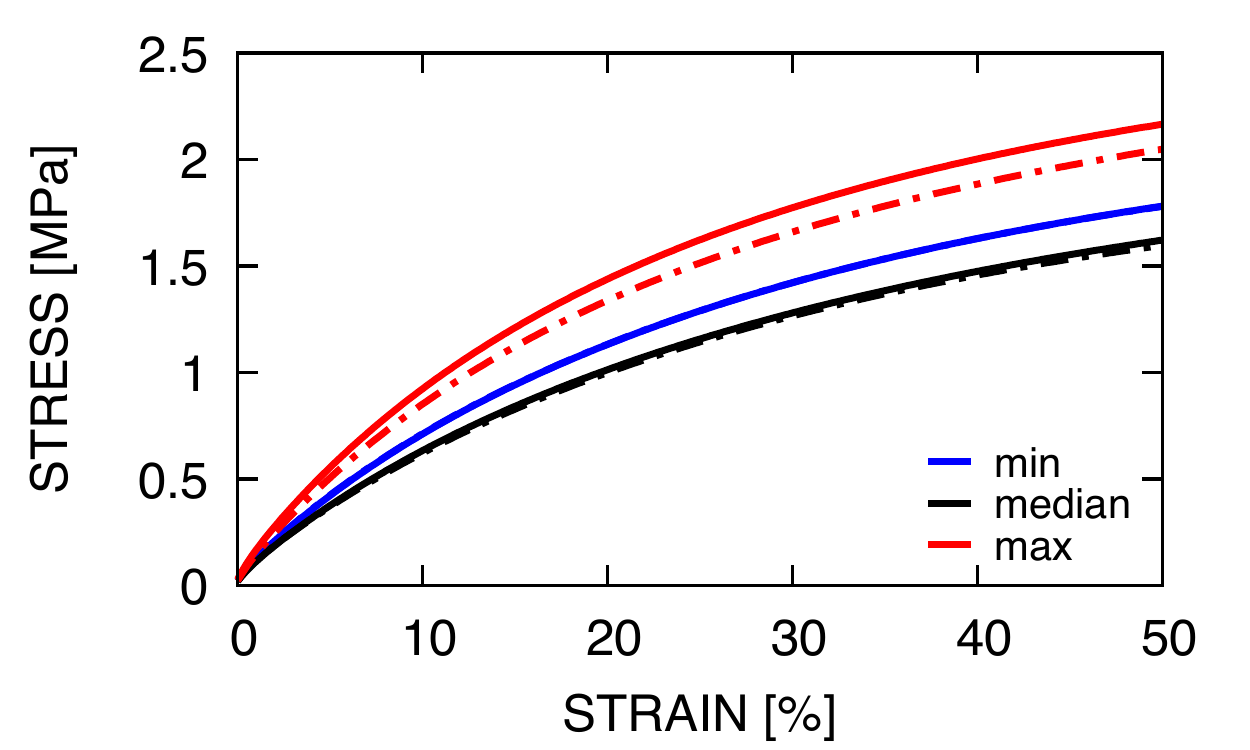}}
\caption{Predictions for response with microstructure.
The predictions (dashed lines) with the minimum (blue), median (black), and maximum (red) RMSE error are show with the true trajectories (solid lines with corresponding colors).
}
\label{fig:gcnn_pred}
\end{figure}

As discussed in \sref{sec:hetero_response} we generated three datasets: (a) porous aluminum with a elastoplastic J2 material model, (b) porous silicone with a viscoelastic UPM material model, and (c) silicone with elastic glass inclusions.
\fref{fig:gcnn_comp} shows the evolution of RMSE of predictions of held-out data over time (note all training/testing data had same duration and time step).
For the porous plastic samples the mean prediction error was 0.0033, and the mean correlation was 0.993 over held-out samples and across time.
The predictions for the porous viscoelastic samples were similarly accurate: the mean prediction error was 0.0037, and the mean correlation was  0.998.
The predictions for the more challenging composite viscoelastic matrix with stiff elastic inclusion samples was slightly less accurate: the mean prediction error was 0.011, and the mean correlation was  0.978.
\fref{fig:gcnn_pred} compares the true and predicted response trajectories for the held out samples with the minimum, median and maximum RMSE.
In fact, even for the worse cases, the predictions (dashed) and true (solid) trajectories are nearly indistinguishable.
Clearly the combined GCNN+\ISVNODE architecture produces accurate predictions that distinguish the effects of varying microstructure and capture their evolution.
This also demonstrates that the GCNN+\ISVNODE is comparably accurate on different classes of inelastic response.
Furthermore, GCNN+\ISVNODE takes 1.47 ms/step to evaluate on a given adjacency matrix and loading history on average; this represents about 3 orders of magnitude speed-up over the traditional model with finite element mesh and the same loading.


\section{Conclusion} \label{sec:conclusion}

By basing a neural ODE-like network on classical ISV theory, the proposed \ISVNODE was able to model a variety of inelastic processes and learn internal states in a general and extensible manner.
Two primary variants, based on how the stress MLP is formulated, were tested.
The potential-based variant allows direct connection with the dissipation requirements of Coleman-Gurtin theory; however, the tensor-basis version has inherent polynomial complexity that reduces the modeling burden on the trainable component functions.
Both are exactly equivariant by design.
This framework precludes the discrepancy/model-form errors that are inherent in calibrating preconceived, traditional models.
In addition we demonstrated the  application of the \ISVNODE to modeling materials with microstructure, and expect that the \ISVNODE will be as effective in modeling materials not easily categorized into elastoplastic or viscoelastic as it was in modeling exemplars of those responses.
Although the state variables themselves do not currently have a precise physical interpretation, the \ISVNODE has components that are analogs of traditional stress and flow rules.
This fact makes \ISVNODE models drop-in replacements for traditional models in existing simulators.
We have demonstrated that accurate NN models can be on the order of 1000$\times$ faster to evaluate than their traditional counterparts.
Furthermore, utilizing a single, general, data-driven constitutive modeling framework will simplify validation of suites of material models needed to simulate full systems.
Also uncertainty quantification (UQ) can be embedded in an \ISVNODE with variational inference-based extensions to the two constituent MLPs \cite{graves2011practical}.

By learning directly from data with \ISVNODE with a tunable set of hidden variables representing material state, the process of building low-discrepancy models should become relatively straightforward.
We expect that this generalized framework for inelastic response will lead to an agile and robust process of constructing constitutive models.
As demonstrated, the state-space can be built with complexity appropriate for the available data and physical process.
The \ISVNODE and its microstructural extension have multiple applications beyond accelerating large-scale simulations such as providing efficient sub-grid models of complex microstructures and surrogate for enabling high-dimensional sampling-based UQ, material optimization, and structure-property maps.

Our findings suggest multiple avenues for future work.
To fully realize the promise of the \ISVNODE framework we intend to extend it to representing non-isothermal processes and embed all the thermodynamic implications of the Coleman-Gurtin ISV theory \cite{coleman1967thermodynamics}.
Since complete sampling of loading modes remains a challenge \cite{fuhg2021physics,logarzo2021smart}, as does incorporating limited experimental data, we are pursuing both active learning \cite{settles2009active,settles2012active} and data fusion techniques \cite{bleiholder2009data,castanedo2013review}.
A related challenge is generating enough microstructure response data to train a model capable to predicting general loading modes.
For this issue we are pursing transfer learning where an \ISVNODE with microstructural inputs is first trained to the homogeneous multimodal data and then to the more limited heterogeneous data.
Lastly, one of the more fundamental challenges results from applying the \ISVNODE framework to elastic-plastic response.
Effectively we are tasking the flow MLP $\NN_\fb$ with learning a vector field, which, in the case of elastoplasticity, is not a smooth one (\ie it is zero in the elastic region and non-zero outside).
In the \ISVNODE formulation proposed in this work the primary input is strain.
With traditional elastoplastic models, like J2, the elastic limits are simpler to describe in stress space, which suggests feeding the predicted stress back into flow MLP.
Other techniques such as basing the flow MLP on a dissipation potential, using a derivative of a single output node as in the stress MLP, may yield improved predictions similar to level set technique used in Vlassis and Sun \cite{vlassis2020sobolev}.

\section*{Acknowledgments}
We are grateful for helpful discussions with Craig Hamel (Sandia), Kevin Long (Sandia) and Coleman Alleman (Sandia).
Sandia National Laboratories is a multimission laboratory managed and operated by National Technology and Engineering Solutions of Sandia, LLC., a wholly owned subsidiary of Honeywell International, Inc., for the U.S. Department of Energy's National Nuclear Security Administration under contract DE-NA-0003525. This paper describes objective technical results and analysis. Any subjective views or opinions that might be expressed in the paper do not necessarily represent the views of the U.S. Department of Energy or the United States Government.


\appendix

\section{A tensor basis for strain and strain rate} \label{app:tb}
\renewcommand{\thesubsection}{\Alph{subsection}}

Based on fundamental developments in tensor function representation theory \cite{boehler1987applications,truesdell2004non}, the stress $\Sb$ as a function of its tensor arguments  has the general representation
\begin{equation} \label{eq:rep}
\Sb = \sum_i \sigma_i \Bb_i \ ,
\end{equation}
where the elements of the basis $\Bb_i \in \Bc$ are known powers of the tensor inputs and the coefficients $\sigma_i = \sigma_i(\Inv)$ are functions of the (scalar) invariants $\Ic$ of the inputs.
If stress is only a function of strain, $\Sb(\Eb)$, its representation consists of three terms
\begin{equation}
\Sb = \sum_{i=0}^2 \sigma_i(\Ic) \Eb^i
\end{equation}
where the basis $\Bb_i$ are powers of strain $\Eb^i$ and $\Ic = \{ \tr \Eb, 1/2(\tr^2 \Eb - \tr \Eb^2), \det \Eb\}$ is a complete set of invariants.
Beyond the fact that both the second Piola-Kirchhof stress $\Sb$ and the Lagrange strain  $\Eb$ are material frame invariant, this representation represents a compact, coordinate-free description.
Note for $\Sb(\Eb)$, $\Sb$ and $\Eb$ are collinear, in that they have the same eigenvectors.

For tensor function of two tensor arguments, such as $\Sb(\Eb,\dot{\Eb})$, Rivlin \cite{rivlin1955further} showed that 10 polynomial invariants
\begin{equation}
\Ic = \{ \tr \Eb, \tr \Eb^2, \tr \Eb^3,
\tr \dot{\Eb}, \tr \dot{\Eb}^2, \tr \dot{\Eb}^3,
\tr \Eb \dot{\Eb}, \tr \Eb \dot{\Eb}^2, \tr \Eb^2 \dot{\Eb},  \tr \Eb^2 \dot{\Eb}^2 \}
\end{equation}
and a 9 element tensor basis
\begin{equation} \label{eq:tb1}
\Bc = \{ \Ib, \Eb, \Eb^2, \dot{\Eb}, \dot{\Eb^2}, \sym \Eb \dot{\Eb}, \sym \Eb^2 \dot{\Eb}, \sym \Eb \dot{\Eb^2}, \sym \Eb^2 \dot{\Eb}^2 \}
\end{equation}
are, in general, necessary.
See also \cite[Ch.3, Eq. 9 and 11]{boehler1987representations} and note here $\dot{\Eb}^2 \equiv \dot{\Eb} \dot{\Eb}$.
However 9 basis elements are more than necessary to represent a symmetric tensor with 6 independent components; hence, the coefficient functions can not be solved for directly (as in \cref{frankel2020tensor}).

For the special case where one argument is the rate of the other can show that only six tensor basis elements:
\begin{equation}
\Bc^* = \{ \Eb^0, \Eb^1, \Eb^2, \dot{\overline{\Eb^1}}, \dot{\overline{\Eb^2}}, \dot{\overline{\Eb^3}} \}
\end{equation}
are required for a complete and irreducible basis.
In addition to reducing the basis this allows for the coefficients of the tensor basis representation to be solved for explicitly.
Note Peters \etal \cite{peters2020s} also employed a reduced basis in a TBNN for fluids.

First recognize that the first three elements of $\Bc^*$, are linearly independent and span the eigenspace of $\Eb$
\begin{equation}
\spn\{ \Eb^p, p=0,1,2 \} = \spn\{ \ab_i \otimes \ab_i, i = 1,2,3 \}
= \spn \Ac
\end{equation}
where
\begin{equation}
\Eb = \sum_i \lambda_i \ab_i \otimes \ab_i
\end{equation}
is the spectral decomposition of $\Eb$, with $\lambda_i$ being its eigenvalues and $\ab_i$ its eigenvectors.
This follows from Wang's Lemma \cite{bowen2008introduction} as outlined by Gurtin \cite[Sec. 37]{gurtin1982introduction} which implies
\begin{equation}
\sum_{j=0}^2 c_j \Eb^j = \mathbf{0}
\ \Leftrightarrow \ \sum_{j=0}^2 c_j \lambda_i^j = 0
\ \Leftrightarrow \ c_j = 0
\end{equation}
for $\Eb$ having three distinct eigenvalues $\lambda_i$, other cases follow by the same logic.

If we now examine the rate
\begin{equation}
\dot{\Eb} = \underbrace{\sum_i \dot{\lambda}_i \ab_i \otimes \ab_i}_{\text{in span of } \Ac}
+  \underbrace{\sum_i \lambda_i \left( \dot{\ab}_i \otimes \ab_i + \ab_i \otimes \dot{\ab}_i \right)}_{\text{ not in span of } \Ac} \ .
\end{equation}
Given the fact that $\dot{\ab}_i \cdot \ab_i = 0$ since $\ab_i$ are unit vectors, $\| \ab_i \| = 1$, it follows that
\begin{equation}
\spn \{ \left( \dot{\ab}_i \otimes \ab_i + \ab_i \otimes \dot{\ab}_i \right)  =
\spn \{ \left( \ab_j \otimes \ab_i + \ab_i \otimes \ab_j \right) , \ i \ne j \ ,
\end{equation}
and these three additional elements complete the basis \cite[Ch.3]{boehler1987representations}.
Unfortunately the eigenvector dyads, $\ab_i \otimes \ab_j$, although being an orthonormal basis, are not a convenient basis since they are expensive to compute and are not permutational invariant like the usual scalar invariants (\eg ordering the eigenvalues does not preserve continuity in time).

Nevertheless if we examine the five remaining basis elements of the general tensor basis:
\begin{equation}
\dot{\Eb} =
\underbrace{\sum_i \dot{\lambda}_i \ab_i \otimes \ab_i}_{\text{in span of } \Ac}
+  \underbrace{\sum_i \lambda_i \left( \dot{\ab}_i \otimes \ab_i + \ab_i \otimes \dot{\ab}_i \right)}_{\text{ not in span of } \Ac}
\end{equation}
\begin{eqnarray}
(\dot{\Eb})^2 &=& \left( \sum_i \dot{\lambda}_i \ab_i \otimes \ab_i
+  \sum_i \lambda_i \left( \dot{\ab}_i \otimes \ab_i + \ab_i \otimes \dot{\ab}_i \right) \right)^2 \\
&=&  \sum_i \underbrace{\dot{\lambda}_i^2 \ab_i \otimes \ab_i}_{\text{in span}}
+  \underbrace{\lambda_i \dot{\lambda}_i \left( \dot{\ab}_i \otimes \ab_i + \ab_i \otimes \dot{\ab}_i\right)}_{\text{not in span}}
+  \underbrace{\lambda_i^2 \left( \dot{\ab}_i \otimes \dot{\ab}_i + (\dot{\ab}_i \cdot \dot{\ab}_i) \ab_i \otimes \ab_i \right)}_{\text{in span}} \nonumber
\end{eqnarray}
\begin{eqnarray}
\sym \Eb \dot{\Eb} &=& \sym \left[
\left(
\sum_i {\lambda}_i \ab_i \otimes \ab_i
\right)
\left(
\sum_i \dot{\lambda}_i \ab_i \otimes \ab_i
+  \sum_i \lambda_i \left( \dot{\ab}_i \otimes \ab_i + \ab_i \otimes \dot{\ab}_i \right) \right)
\right] \nonumber \\
&=&
\underbrace{\sum_i \lambda_i \dot{\lambda}_i \ab_i \otimes \ab_i}_{\text{in span}}
+  \underbrace{\sum_i \lambda_i^2 \left( \dot{\ab}_i \otimes \ab_i + \ab_i \otimes \dot{\ab}_i \right)}_{\text{ not in span } }
\end{eqnarray}
\begin{eqnarray}
\sym \Eb^2 \dot{\Eb}
&=&
\underbrace{\sum_i \lambda_i^2 \dot{\lambda}_i \ab_i \otimes \ab_i}_{\text{in span } }
+  \underbrace{\sum_i \lambda_i^3 \left( \dot{\ab}_i \otimes \ab_i + \ab_i \otimes \dot{\ab}_i \right)}_{\text{ not in span } }
\end{eqnarray}
\begin{eqnarray}
\sym \Eb \dot{\Eb}^2 &=& \sym \Biggl[
\left(
\sum_i {\lambda}_i \ab_i \otimes \ab_i
\right) \\
&& \left(
\sum_i
\dot{\lambda}_i^2 \ab_i \otimes \ab_i
+ \lambda_i \dot{\lambda}_i \left( \dot{\ab}_i \otimes \ab_i + \ab_i \otimes \dot{\ab}_i \right)
+ \lambda_i^2  \left( \dot{\ab}_i \otimes \dot{\ab}_i + (\dot{\ab}_i \cdot \dot{\ab}_i) \ab_i \otimes \ab_i \right)
\right)
\Biggr] \nonumber \\
&=&
\sum_i \underbrace{\lambda_i \dot{\lambda}_i^2 \ab_i \otimes \ab_i}_{\text{in span}}
+ \underbrace{ \lambda_i^2 \dot{\lambda}_i \left( \dot{\ab}_i \otimes {\ab}_i + {\ab}_i \otimes \dot{\ab}_i) \right)}_{\text{not in span}}
+ \underbrace{ \lambda_i^3 \left( \dot{\ab}_i \otimes \dot{\ab}_i + (\dot{\ab}_i \cdot \dot{\ab}_i) \ab_i \otimes \ab_i \right)}_{\text{in span}} \nonumber
\end{eqnarray}
we observe that the parts of $\{ \dot{\Eb}, \sym \Eb \dot{\Eb}, \sym \Eb \dot{\Eb} \}$ not in the span of $\Ac$ depend on $\lambda_i$, $\lambda_i^2$, and $\lambda_i^3$, respectively,
Hence, Wang's Lemma can be used again (with the trivial restriction $\lambda_i > 0$) to state
\begin{equation}
\spn \{ \dot{\Eb}, \sym \Eb \dot{\Eb}, \sym \Eb \dot{\Eb} \} =
\spn \{ \left( \dot{\ab}_i \otimes \ab_i + \ab_i \otimes \dot{\ab}_i \right), \ 0 \le i < 3 \} \ .
\end{equation}
Furthermore, since $\dot{\overline{\Eb^a}} = a \Eb^{a-1} \dot{\Eb}$ we can claim
\begin{equation}
\spn \{ \dot{\overline{\Eb}^a}, \ 0 < a \le 3 \} =
\spn \{ \dot{\Eb}, \sym \Eb \dot{\Eb}, \sym \Eb^2 \dot{\Eb} \}
\end{equation}
and finally
$\Bc^* = \{ \Eb^a \ | \ 0 \le a < 3 \} \ \cup \ \{ \dot{\overline{\Eb^a}} \ | \ 0 < a \le 3 \}$ has the same span as $\Bc$ and is linearly independent.

Beyond reducing the complexity of the representation, this reduced basis enables solving for the coefficient functions directly as in \cref{frankel2020tensor}.
Although we do not employ the explicit method of solving for the coefficient functions directly in this work it has advantages in training TBNNs.
It does, however, have a complication when there is multiplicity in the eigenvalues of the input tensors.
Although in finite precision arithmetic this is rarely encountered, it does degrade the conditioning of the linear system that must be solved.
Similarly simple loadings were $\dot{\Eb}$ and $\Eb$ are collinear and/or $\dot{\ab}_i = \mathbf{0}$ can lead to rank deficiencies.
Gurtin \cite{gurtin1982introduction} provided a well-conditioned solution by way of solving a  reduced system.
For instance, in the case of $\Sb=\Sb(\Eb)$ and two of the eigenvalues of $\Eb$ are identical (as in uniaxial tension), $\Eb$ can be represented as $\Eb = \lambda_1 \ab\otimes\ab + \lambda_2 (\Ib - \ab\otimes\ab)$ where $\lambda_1$ is the unique eigenvalue and $\lambda_2$ is the repeated one.
Now instead of solving
\begin{equation} \label{eq:system}
\begin{bmatrix}
\varsigma_1 \\ \varsigma_2 \\ \varsigma_3
\end{bmatrix}
=
\begin{bmatrix}
1 & \lambda_1 & \lambda_1^2\\
1 & \lambda_2 & \lambda_2^2\\
1 & \lambda_3 & \lambda_3^2\\
\end{bmatrix}
\begin{bmatrix}
\sigma_0 \\ \sigma_1 \\ \sigma_2
\end{bmatrix}.
\end{equation}
where $\varsigma_i$ are the eigenvalues of $\Sb$,
only
\begin{equation} \label{eq:system2}
\begin{bmatrix}
\varsigma_1 \\ \varsigma_2
\end{bmatrix}
=
\begin{bmatrix}
1 & \lambda_1 \\
1 & \lambda_2 \\
\end{bmatrix}
\begin{bmatrix}
\sigma_0 \\ \sigma_1
\end{bmatrix}.
\end{equation}
needs to be solved since the coefficient $\sigma_2 = 0$.
We proposed an alternative scheme in \cref{frankel2020tensor}.


\begin{thebibliography}{10}

\bibitem{coleman1961foundations}
Bernard~D Coleman and Walter Noll.
\newblock Foundations of linear viscoelasticity.
\newblock {\em Reviews of modern physics}, 33(2):239, 1961.

\bibitem{lubliner1969fading}
J~Lubliner.
\newblock On fading memory in materials of evolutionary type.
\newblock {\em Acta Mechanica}, 8(1):75--81, 1969.

\bibitem{coleman1967thermodynamics}
Bernard~D Coleman and Morton~E Gurtin.
\newblock Thermodynamics with internal state variables.
\newblock {\em The Journal of Chemical Physics}, 47(2):597--613, 1967.

\bibitem{kratochvil1969thermodynamics}
J~Kratochvil and OW~Dillon~Jr.
\newblock Thermodynamics of elastic-plastic materials as a theory with internal
state variables.
\newblock {\em Journal of Applied Physics}, 40(8):3207--3218, 1969.

\bibitem{chorin2000optimal}
Alexandre~J Chorin, Ole~H Hald, and Raz Kupferman.
\newblock Optimal prediction and the {M}ori--{Z}wanzig representation of
irreversible processes.
\newblock {\em Proceedings of the National Academy of Sciences},
97(7):2968--2973, 2000.

\bibitem{li2014construction}
Zhen Li, Xin Bian, Bruce Caswell, and George~Em Karniadakis.
\newblock Construction of dissipative particle dynamics models for complex
fluids via the {M}ori--{Z}wanzig formulation.
\newblock {\em Soft Matter}, 10(43):8659--8672, 2014.

\bibitem{parish2017dynamic}
Eric~J Parish and Karthik Duraisamy.
\newblock A dynamic subgrid scale model for large eddy simulations based on the
{M}ori--{Z}wanzig formalism.
\newblock {\em Journal of Computational Physics}, 349:154--175, 2017.

\bibitem{adelman1976generalized}
SA~Adelman and JD~Doll.
\newblock Generalized langevin equation approach for atom/solid-surface
scattering: General formulation for classical scattering off harmonic solids.
\newblock {\em The Journal of chemical physics}, 64(6):2375--2388, 1976.

\bibitem{wagner2003coupling}
Gregory~J Wagner and Wing~Kam Liu.
\newblock Coupling of atomistic and continuum simulations using a bridging
scale decomposition.
\newblock {\em Journal of Computational Physics}, 190(1):249--274, 2003.

\bibitem{mcdowell2005internal}
DL~McDowell.
\newblock Internal state variable theory.
\newblock In {\em Handbook of Materials Modeling}, pages 1151--1169. Springer,
2005.

\bibitem{horstemeyer2010historical}
Mark~F Horstemeyer and Douglas~J Bammann.
\newblock Historical review of internal state variable theory for inelasticity.
\newblock {\em International Journal of Plasticity}, 26(9):1310--1334, 2010.

\bibitem{truesdell2004non}
Clifford Truesdell and Walter Noll.
\newblock The non-linear field theories of mechanics.
\newblock In {\em The non-linear field theories of mechanics}, pages 1--579.
Springer, 2004.

\bibitem{kroner1965how}
E~Kr{\"o}ner.
\newblock How the internal state of a plastically deformed body is to be
described in a continuum theory.
\newblock In {\em Proceedings of the Fourth International Congress on
Rheology}, pages c32 N67--14522. Wiley, 1965.

\bibitem{rice1971inelastic}
James~R Rice.
\newblock Inelastic constitutive relations for solids: an internal-variable
theory and its application to metal plasticity.
\newblock {\em Journal of the Mechanics and Physics of Solids}, 19(6):433--455,
1971.

\bibitem{bhandari1973unified}
DR~Bhandari and JT~Oden.
\newblock A unified theory of thermoviscoplasticity of crystalline solids.
\newblock {\em International Journal of Non-Linear Mechanics}, 8(3):261--277,
1973.

\bibitem{perzyna1986internal}
Piotr Perzyna.
\newblock Internal state variable description of dynamic fracture of ductile
solids.
\newblock {\em International Journal of Solids and Structures}, 22(7):797--818,
1986.

\bibitem{reese1998theory}
Stefanie Reese and Sanjay Govindjee.
\newblock A theory of finite viscoelasticity and numerical aspects.
\newblock {\em International journal of solids and structures},
35(26-27):3455--3482, 1998.

\bibitem{simo1992associative}
JC~Simo and Ch~Miehe.
\newblock Associative coupled thermoplasticity at finite strains: Formulation,
numerical analysis and implementation.
\newblock {\em Computer Methods in Applied Mechanics and Engineering},
98(1):41--104, 1992.

\bibitem{simo1992algorithms}
Juan~C Simo.
\newblock Algorithms for static and dynamic multiplicative plasticity that
preserve the classical return mapping schemes of the infinitesimal theory.
\newblock {\em Computer Methods in Applied Mechanics and Engineering},
99(1):61--112, 1992.

\bibitem{germain1983continuum}
Paul Germain, Pierre Suquet, and Quoc~Son Nguyen.
\newblock Continuum thermodynamics.
\newblock {\em ASME Transactions Series E Journal of Applied Mechanics},
50:1010--1020, 1983.

\bibitem{onsager1931reciprocalI}
Lars Onsager.
\newblock Reciprocal relations in irreversible processes. {I}.
\newblock {\em Physical review}, 37(4):405, 1931.

\bibitem{onsager1931reciprocalII}
Lars Onsager.
\newblock Reciprocal relations in irreversible processes. {II}.
\newblock {\em Physical review}, 38(12):2265, 1931.

\bibitem{eckart1940thermodynamics}
Carl Eckart.
\newblock The thermodynamics of irreversible processes. {I}. {T}he simple
fluid.
\newblock {\em Physical Review}, 58(3):267, 1940.

\bibitem{eckart1948thermodynamics}
Carl Eckart.
\newblock The thermodynamics of irreversible processes. {IV}. {T}he theory of
elasticity and anelasticity.
\newblock {\em Physical Review}, 73(4):373, 1948.

\bibitem{ling2016machine}
Julia Ling, Reese Jones, and Jeremy Templeton.
\newblock Machine learning strategies for systems with invariance properties.
\newblock {\em Journal of Computational Physics}, 318:22--35, 2016.

\bibitem{raissi2019physics}
Maziar Raissi, Paris Perdikaris, and George~E Karniadakis.
\newblock Physics-informed neural networks: A deep learning framework for
solving forward and inverse problems involving nonlinear partial differential
equations.
\newblock {\em Journal of Computational Physics}, 378:686--707, 2019.

\bibitem{lee2019deep}
Kookjin Lee and Kevin Carlberg.
\newblock Deep conservation: A latent dynamics model for exact satisfaction of
physical conservation laws.
\newblock {\em arXiv preprint arXiv:1909.09754}, 2019.

\bibitem{wang2020incorporating}
Rui Wang, Robin Walters, and Rose Yu.
\newblock Incorporating symmetry into deep dynamics models for improved
generalization.
\newblock {\em arXiv preprint arXiv:2002.03061}, 2020.

\bibitem{karniadakis2021physics}
George~Em Karniadakis, Ioannis~G Kevrekidis, Lu~Lu, Paris Perdikaris, Sifan
Wang, and Liu Yang.
\newblock Physics-informed machine learning.
\newblock {\em Nature Reviews Physics}, 3(6):422--440, 2021.

\bibitem{linka2021constitutive}
Kevin Linka, Markus Hillg{\"a}rtner, Kian~P Abdolazizi, Roland~C Aydin, Mikhail
Itskov, and Christian~J Cyron.
\newblock Constitutive artificial neural networks: A fast and general approach
to predictive data-driven constitutive modeling by deep learning.
\newblock {\em Journal of Computational Physics}, 429:110010, 2021.

\bibitem{masi2021thermodynamics}
Filippo Masi, Ioannis Stefanou, Paolo Vannucci, and Victor Maffi-Berthier.
\newblock Thermodynamics-based artificial neural networks for constitutive
modeling.
\newblock {\em Journal of the Mechanics and Physics of Solids}, 147:104277,
2021.

\bibitem{tsoi1997discrete}
Ah~Chung Tsoi and Andrew Back.
\newblock Discrete time recurrent neural network architectures: A unifying
review.
\newblock {\em Neurocomputing}, 15(3-4):183--223, 1997.

\bibitem{yu2019review}
Yong Yu, Xiaosheng Si, Changhua Hu, and Jianxun Zhang.
\newblock A review of recurrent neural networks: Lstm cells and network
architectures.
\newblock {\em Neural computation}, 31(7):1235--1270, 2019.

\bibitem{lipton2015critical}
Zachary~C Lipton, John Berkowitz, and Charles Elkan.
\newblock A critical review of recurrent neural networks for sequence learning.
\newblock {\em arXiv preprint arXiv:1506.00019}, 2015.

\bibitem{kobyzev2020normalizing}
Ivan Kobyzev, Simon Prince, and Marcus Brubaker.
\newblock Normalizing flows: An introduction and review of current methods.
\newblock {\em IEEE Transactions on Pattern Analysis and Machine Intelligence},
2020.

\bibitem{hochreiter1997long}
Sepp Hochreiter and J{\"u}rgen Schmidhuber.
\newblock Long short-term memory.
\newblock {\em Neural computation}, 9(8):1735--1780, 1997.

\bibitem{cho2014learning}
Kyunghyun Cho, Bart Van~Merri{\"e}nboer, Caglar Gulcehre, Dzmitry Bahdanau,
Fethi Bougares, Holger Schwenk, and Yoshua Bengio.
\newblock Learning phrase representations using {RNN} encoder-decoder for
statistical machine translation.
\newblock {\em arXiv preprint arXiv:1406.1078}, 2014.

\bibitem{he2016deep}
Kaiming He, Xiangyu Zhang, Shaoqing Ren, and Jian Sun.
\newblock Deep residual learning for image recognition.
\newblock In {\em Proceedings of the IEEE conference on computer vision and
pattern recognition}, pages 770--778, 2016.

\bibitem{chen2018neural}
Ricky~TQ Chen, Yulia Rubanova, Jesse Bettencourt, and David Duvenaud.
\newblock Neural ordinary differential equations.
\newblock {\em arXiv preprint arXiv:1806.07366}, 2018.

\bibitem{dupont2019augmented}
Emilien Dupont, Arnaud Doucet, and Yee~Whye Teh.
\newblock Augmented neural odes.
\newblock {\em arXiv preprint arXiv:1904.01681}, 2019.

\bibitem{coddington1955theory}
Earl~A Coddington and Norman Levinson.
\newblock {\em Theory of ordinary differential equations}.
\newblock Tata McGraw-Hill Education, 1955.

\bibitem{arnold1973ordinary}
Vladimir~Igorevic Arnold.
\newblock {\em Ordinary differential equations}.
\newblock MIT press, 1973.

\bibitem{rackauckas2020universal}
Christopher Rackauckas, Yingbo Ma, Julius Martensen, Collin Warner, Kirill
Zubov, Rohit Supekar, Dominic Skinner, Ali Ramadhan, and Alan Edelman.
\newblock Universal differential equations for scientific machine learning.
\newblock {\em arXiv preprint arXiv:2001.04385}, 2020.

\bibitem{rubel1981universal}
Lee~A Rubel et~al.
\newblock A universal differential equation.
\newblock {\em Bulletin (new series) of the american mathematical society},
4(3):345--349, 1981.

\bibitem{lu2019deeponet}
Lu~Lu, Pengzhan Jin, and George~Em Karniadakis.
\newblock Deeponet: Learning nonlinear operators for identifying differential
equations based on the universal approximation theorem of operators.
\newblock {\em arXiv preprint arXiv:1910.03193}, 2019.

\bibitem{teshima2020universal}
Takeshi Teshima, Koichi Tojo, Masahiro Ikeda, Isao Ishikawa, and Kenta Oono.
\newblock Universal approximation property of neural ordinary differential
equations.
\newblock {\em arXiv preprint arXiv:2012.02414}, 2020.

\bibitem{hornik1989multilayer}
Kurt Hornik, Maxwell Stinchcombe, and Halbert White.
\newblock Multilayer feedforward networks are universal approximators.
\newblock {\em Neural networks}, 2(5):359--366, 1989.

\bibitem{scarselli1998universal}
Franco Scarselli and Ah~Chung Tsoi.
\newblock Universal approximation using feedforward neural networks: A survey
of some existing methods, and some new results.
\newblock {\em Neural networks}, 11(1):15--37, 1998.

\bibitem{dandekar2020bayesian}
Raj Dandekar, Karen Chung, Vaibhav Dixit, Mohamed Tarek, Aslan Garcia-Valadez,
Krishna~Vishal Vemula, and Chris Rackauckas.
\newblock Bayesian neural ordinary differential equations.
\newblock {\em arXiv preprint arXiv:2012.07244}, 2020.

\bibitem{fu2020learning}
Xiaohan Fu, Lo-Bin Chang, and Dongbin Xiu.
\newblock Learning reduced systems via deep neural networks with memory.
\newblock {\em Journal of Machine Learning for Modeling and Computing}, 1(2),
2020.

\bibitem{qin2021data}
Tong Qin, Zhen Chen, John~D Jakeman, and Dongbin Xiu.
\newblock Data-driven learning of nonautonomous systems.
\newblock {\em SIAM Journal on Scientific Computing}, 43(3):A1607--A1624, 2021.

\bibitem{drgona2020spectral}
Jan Drgona, Elliott Skomski, Soumya Vasisht, Aaron Tuor, and Draguna Vrabie.
\newblock Spectral analysis and stability of deep neural dynamics.
\newblock {\em arXiv preprint arXiv:2011.13492}, 2020.

\bibitem{lagaris1998artificial}
Isaac~E Lagaris, Aristidis Likas, and Dimitrios~I Fotiadis.
\newblock Artificial neural networks for solving ordinary and partial
differential equations.
\newblock {\em IEEE transactions on neural networks}, 9(5):987--1000, 1998.

\bibitem{jones2018machine}
RE~Jones, JA~Templeton, CM~Sanders, and JT~Ostien.
\newblock Machine learning models of plastic flow based on representation
theory.
\newblock {\em Computer Modeling in Engineering \& Sciences}, 117(3):309--342,
2018.

\bibitem{frankel2020tensor}
Ari~L Frankel, Reese~E Jones, and Laura~P Swiler.
\newblock Tensor basis gaussian process models of hyperelastic materials.
\newblock {\em Journal of Machine Learning for Modeling and Computing}, 1(1),
2020.

\bibitem{peters2020s}
Eric~L Peters, Riccardo Balin, Kenneth~E Jansen, Alireza Doostan, and John~A
Evans.
\newblock S-frame discrepancy correction models for data-informed {R}eynolds
stress closure.
\newblock {\em arXiv preprint arXiv:2004.08865}, 2020.

\bibitem{xu2020inverse}
Kailai Xu, Alexandre~M Tartakovsky, Jeff Burghardt, and Eric Darve.
\newblock Inverse modeling of viscoelasticity materials using physics
constrained learning.
\newblock {\em arXiv preprint arXiv:2005.04384}, 2020.

\bibitem{logarzo2021smart}
Hernan~J Logarzo, German Capuano, and Julian~J Rimoli.
\newblock Smart constitutive laws: Inelastic homogenization through machine
learning.
\newblock {\em Computer Methods in Applied Mechanics and Engineering},
373:113482, 2021.

\bibitem{vlassis2020sobolev}
Nikolaos~N Vlassis and WaiChing Sun.
\newblock Sobolev training of thermodynamic-informed neural networks for
smoothed elasto-plasticity models with level set hardening.
\newblock {\em arXiv preprint arXiv:2010.11265}, 2020.

\bibitem{teichert2019machine}
Gregory~H Teichert, AR~Natarajan, A~Van~der Ven, and Krishna Garikipati.
\newblock Machine learning materials physics: Integrable deep neural networks
enable scale bridging by learning free energy functions.
\newblock {\em Computer Methods in Applied Mechanics and Engineering},
353:201--216, 2019.

\bibitem{frankel2019predicting}
Ari~L Frankel, Reese~E Jones, Coleman Alleman, and Jeremy~A Templeton.
\newblock Predicting the mechanical response of oligocrystals with deep
learning.
\newblock {\em Computational Materials Science}, 169:109099, 2019.

\bibitem{frankel2020prediction}
Ari Frankel, Kousuke Tachida, and Reese Jones.
\newblock Prediction of the evolution of the stress field of polycrystals
undergoing elastic-plastic deformation with a hybrid neural network model.
\newblock {\em Machine Learning: Science and Technology}, 1(3):035005, 2020.

\bibitem{vlassis2020geometric}
Nikolaos~N Vlassis, Ran Ma, and WaiChing Sun.
\newblock Geometric deep learning for computational mechanics part {I}:
Anisotropic hyperelasticity.
\newblock {\em Computer Methods in Applied Mechanics and Engineering},
371:113299, 2020.

\bibitem{frankel2021mesh}
Ari Frankel, Cosmin Safta, Coleman Alleman, and Reese Jones.
\newblock Mesh-based graph convolutional neural networks for modeling materials
with microstructure.
\newblock {\em Journal of Machine Learning for Modeling and Computation}, 2021.

\bibitem{simo2006computational}
Juan~C Simo and Thomas~JR Hughes.
\newblock {\em Computational inelasticity}, volume~7.
\newblock Springer Science \& Business Media, 2006.

\bibitem{gurtin2010mechanics}
Morton~E Gurtin, Eliot Fried, and Lallit Anand.
\newblock {\em The mechanics and thermodynamics of continua}.
\newblock Cambridge University Press, 2010.

\bibitem{silhavy2013mechanics}
Miroslav Silhavy.
\newblock {\em The mechanics and thermodynamics of continuous media}.
\newblock Springer Science \& Business Media, 2013.

\bibitem{truesdell1959rational}
C~Truesdell.
\newblock The rational mechanics of materials-past, present, future.
\newblock {\em Appl. Mech. Reviews}, 12:75--80, 1959.

\bibitem{truesdell1960classical}
Clifford Truesdell and Richard Toupin.
\newblock The classical field theories.
\newblock In {\em Principles of classical mechanics and field theory/Prinzipien
der Klassischen Mechanik und Feldtheorie}, pages 226--858. Springer, 1960.

\bibitem{haupt1989application}
P~Haupt and Ch~Tsakmakis.
\newblock On the application of dual variables in continuum mechanics.
\newblock {\em Continuum Mechanics and Thermodynamics}, 1(3):165--196, 1989.

\bibitem{rosenblatt1961principles}
Frank Rosenblatt.
\newblock Principles of neurodynamics. perceptrons and the theory of brain
mechanisms.
\newblock Technical report, Cornell Aeronautical Lab Inc Buffalo NY, 1961.

\bibitem{goodfellow2016deep}
Ian Goodfellow, Yoshua Bengio, Aaron Courville, and Yoshua Bengio.
\newblock {\em Deep learning}.
\newblock MIT press Cambridge, 2016.

\bibitem{rumelhart1986learning}
David~E Rumelhart, Geoffrey~E Hinton, and Ronald~J Williams.
\newblock Learning representations by back-propagating errors.
\newblock {\em nature}, 323(6088):533--536, 1986.

\bibitem{robbins1951stochastic}
Herbert Robbins and Sutton Monro.
\newblock A stochastic approximation method.
\newblock {\em The annals of mathematical statistics}, pages 400--407, 1951.

\bibitem{tensorflow}
{TensorFlow}: An end-to-end open source machine learning platform.

\bibitem{spektral}
Daniele Grattarola.
\newblock {Spektral}: a python library for graph deep learning.
\newblock https://graphneural.network/, 2021.

\bibitem{kingma2014adam}
Diederik~P Kingma and Jimmy Ba.
\newblock Adam: A method for stochastic optimization.
\newblock {\em arXiv preprint arXiv:1412.6980}, 2014.

\bibitem{gunther2020layer}
Stefanie Gunther, Lars Ruthotto, Jacob~B Schroder, Eric~C Cyr, and Nicolas~R
Gauger.
\newblock Layer-parallel training of deep residual neural networks.
\newblock {\em SIAM Journal on Mathematics of Data Science}, 2(1):1--23, 2020.

\bibitem{sierra}
James~R Stewart, H~Carter Edwards, and et~al.
\newblock Sierra mechanics.
\newblock https://www.sandia.gov/ASC/integrated\_codes.html, 2020.

\bibitem{adolf2009simplified}
Douglas~B Adolf, Robert~S Chambers, and Matthew~A Neidigk.
\newblock A simplified potential energy clock model for glassy polymers.
\newblock {\em Polymer}, 50(17):4257--4269, 2009.

\bibitem{long2017linear}
Kevin~Nicholas Long and Judith~Alice Brown.
\newblock A linear viscoelastic model calibration of sylgard 184.
\newblock Technical report, Sandia National Lab.(SNL-NM), Albuquerque, NM
(United States), 2017.
\newblock https://www.osti.gov/biblio/1365535.

\bibitem{lubliner2008plasticity}
Jacob Lubliner.
\newblock {\em Plasticity theory}.
\newblock Courier Corporation, 2008.

\bibitem{brown2018multiscale}
JA~Brown, JD~Carroll, B~Huddleston, Z~Casias, and KN~Long.
\newblock A multiscale study of damage in elastomeric syntactic foams.
\newblock {\em Journal of materials science}, 53(14):10479--10498, 2018.

\bibitem{owen2017hexahedral}
Steven~J Owen, Judith~A Brown, Corey~D Ernst, Hojun Lim, and Kevin~N Long.
\newblock Hexahedral mesh generation for computational materials modeling.
\newblock {\em Procedia engineering}, 203:167--179, 2017.

\bibitem{owen2014parallel}
Steven~J Owen, Matthew~L Staten, and Marguerite~C Sorensen.
\newblock Parallel hexahedral meshing from volume fractions.
\newblock {\em Engineering with Computers}, 30(3):301--313, 2014.

\bibitem{graves2011practical}
Alex Graves.
\newblock Practical variational inference for neural networks.
\newblock {\em Advances in neural information processing systems}, 24, 2011.

\bibitem{fuhg2021physics}
Jan~Niklas Fuhg and Nikolaos Bouklas.
\newblock On physics-informed data-driven isotropic and anisotropic
constitutive models through probabilistic machine learning and space-filling
sampling.
\newblock {\em arXiv preprint arXiv:2109.11028}, 2021.

\bibitem{settles2009active}
Burr Settles.
\newblock Active learning literature survey.
\newblock Technical report, University of Wisconsin-Madison Department of
Computer Sciences, 2009.
\newblock http://burrsettles.com/pub/settles.activelearning.pdf.

\bibitem{settles2012active}
Burr Settles.
\newblock {\em Active learning: Synthesis lectures on artificial intelligence
and machine learning}.
\newblock Long Island, NY: Morgan \& Clay Pool, 2012.

\bibitem{bleiholder2009data}
Jens Bleiholder and Felix Naumann.
\newblock Data fusion.
\newblock {\em ACM computing surveys (CSUR)}, 41(1):1--41, 2009.

\bibitem{castanedo2013review}
Federico Castanedo.
\newblock A review of data fusion techniques.
\newblock {\em The scientific world journal}, 2013, 2013.

\bibitem{boehler1987applications}
Jean-Paul Boehler and Jean-Paul Boehler.
\newblock {\em Applications of tensor functions in solid mechanics}, volume
292.
\newblock Springer, 1987.

\bibitem{rivlin1955further}
Ronald~S. Rivlin.
\newblock Further remarks on the stress-deformation relations for isotropic
materials.
\newblock {\em Journal of Rational Mechanics and Analysis}, 4:681--702, 1955.

\bibitem{boehler1987representations}
JP~Boehler.
\newblock Representations for isotropic and anisotropic non-polynomial tensor
functions.
\newblock In {\em Applications of tensor functions in solid mechanics}, pages
31--53. Springer, 1987.

\bibitem{bowen2008introduction}
Ray~M Bowen and Chao-Cheng Wang.
\newblock {\em Introduction to vectors and tensors}, volume~2.
\newblock Courier Corporation, 2008.

\bibitem{gurtin1982introduction}
Morton~E Gurtin.
\newblock {\em An introduction to continuum mechanics}.
\newblock Academic press, 1982.

\end{thebibliography}
\end{document}